\documentclass[11pt,a4paper]{article}

\usepackage[T1]{fontenc}
\usepackage[utf8]{inputenc}
\usepackage[margin=2.5cm]{geometry}
\usepackage{amsmath,amssymb,amsthm}
\usepackage{graphicx}
\usepackage{booktabs}
\usepackage{microtype}
\usepackage[numbers,square,sort&compress]{natbib}
\usepackage{url}
\usepackage[colorlinks=true,linkcolor=blue,citecolor=blue,urlcolor=blue]{hyperref}

\title{Higher-Order Interactions in Complex Systems:\\
Mechanisms, Behaviour, Representation and Reducibility}

\author{Francisco J. P\'erez-Reche\thanks{%
School of Natural and Computing Sciences, University of Aberdeen,
Aberdeen AB24 3UE, United Kingdom.
\texttt{fperez-reche@abdn.ac.uk}}}

\date{\today}

\begin{document}

\maketitle

\begin{abstract}
Higher-order interactions couple three or more elements in ways not decomposable into pairwise contributions. They are associated with phenomenological signatures, collective thermodynamic or dynamical properties such as explosive synchronization and discontinuous transitions. Neither such signatures nor a many-body description establishes a many-body mechanism, since both depend on how the system is represented. Connecting work on higher-order network dynamics and multivariate statistical dependence, this review distinguishes the mechanism governing a system, the behaviour it produces and the representation in which either is expressed. Behaviour comprises phenomenological signatures and statistical dependence. Representation covers which variables are included and how relations are encoded. We examine direct many-body mechanisms, nonlinear but pairwise mechanisms producing similar phenomenological signatures, indirect routes through shared environments or eliminated variables, and temporal aggregation, an artefact of observation. The same behaviour can arise from different mechanisms, and a higher-order mechanism has no universal behavioural consequence. We formulate reducibility, whether a higher-order model can be replaced by a lower-order one, relative to an admissible model class, a target and a tolerance, and apply it across synchronization, ecology, statistical mechanics and contagion. Information-theoretic and inferential methods are reviewed for detecting higher-order dependence, testing lower-order alternatives and reconstructing interaction structure. What data can establish about the generating model is a question of identifiability, distinct from reducibility. Observational data alone cannot uniquely establish the underlying mechanism without additional structural assumptions, sufficiently resolved dynamics, or interventions. We close with open problems in linking mechanisms to measurable signatures, coarse-graining, identifiability and causality.
\end{abstract}


\vspace{1em}

\section{Introduction}
\label{sec:intro}
%%% ============================================================

Complex systems have traditionally been modelled as networks of pairwise
interactions, in which the fundamental object is the dyadic link between two
elements.
 In recent years, however, a rapidly growing literature has argued that many systems are shaped by \emph{higher-order} interactions that couple three or more elements at once and cannot be reduced to a superposition of pairwise links~\citep{Battiston2020,%
Battiston2021,Bick2023,MajhiPercGhosh2022,Boccaletti2023}. Examples of such systems include neurons, ecological communities, social groups, and coupled oscillators. Mathematical representations such as hypergraphs and simplicial complexes have become dominant tools for capturing these interactions, and a large body of work has developed the structure, dynamics, and applications of these ``higher-order networks.''

Much of the interest in higher-order interactions is motivated by observed phenomena: higher-order models can produce explosive synchronization~\citep{SkardalArenas2020}, bistable epidemic thresholds~\citep{Iacopini2019}, and altered ecological stability~\citep{Grilli2017}. Such system behaviours have often been taken as evidence that higher-order structure is necessary.

Higher-order interactions are not new. $p$-spin models long showed how interactions between groups of $p>2$ spins alter glass transitions~\citep{Derrida1980,GrossMezard1984}, ecologists have debated since the late 1960s whether a third species can modify the interaction between two others in ways a pairwise description misses~\citep{BillickCase1994}, multiplayer games have used group-dependent payoffs for decades~\citep{Gokhale2010}, and neural population studies have debated pairwise versus higher-order descriptions since the mid-2000s~\citep{Schneidman2006}. These traditions developed largely independently, with different formalisms and different criteria for what counts as a higher-order effect.

A key question is whether higher-order interactions are actually necessary to explain collective behaviour, and how that necessity can be established. The difficulty is a confusion of three distinct questions: what mechanism governs the system, what behaviour that mechanism produces, and what representation is used to encode either. A hypergraph may offer a convenient representation without the underlying dynamics being genuinely higher-order~\citep{Peixoto2026,Rommens2026}. A behavioural feature such as synergy or a discontinuous transition can arise from distinct mechanisms, so observing it does not by itself identify which mechanism produced it~\citep{Rosas2022,Marinazzo2025,Robiglio2025}. And apparent group dependence can emerge indirectly, through hidden variables or a shared environment~\citep{Env2026}, or from the time resolution at which the system is observed~\citep{Cencetti2021}, with no direct many-body interaction at all. Keeping these three questions apart, rather than letting a single word such as ``reducible'' settle all of them at once, is the organizing task of what follows.

Our aim is to provide a critical synthesis rather than an encyclopedic survey.
Several of the distinctions involved are recognized individually in the
literature. The contribution of this review is to bring them together into a single framework that separates mechanism, behaviour and representation, and to use it to assess when higher-order descriptions are necessary. We connect studies of
higher-order network dynamics~\citep{Battiston2020,Battiston2021,Battiston2026} with
approaches to multivariate statistical dependence~\citep{Rosas2019,Rosas2022},
distinguishing mechanistic interactions, statistical dependence and phenomenological
signatures, together with the representation in which each is expressed. Across
deterministic, stochastic and inferential settings, we examine whether a lower-order
model can reproduce a specified target, under explicit restrictions on the admissible
models and the required accuracy. This provides a common basis for comparing results
that address different notions of
reducibility~\citep{Neuhauser2020,Llabres2026,Lucas2026,Xie2026,Rommens2026b}, and for
distinguishing successful model reduction from limitations of mechanistic inference.

The review is organized by modelling approach. Section~\ref{sec:reps} fixes the terminology, distinguishing mechanism, behaviour and representation, separating the direct, indirect and nonlinear but pairwise routes to higher-order dependence and phenomenological signatures, and setting out the reduction question and its three subquestions, which are formalized in Section~\ref{sec:synthesis}. Sections~\ref{sec:deterministic} and~\ref{sec:stochastic} are about models of systems with direct interaction mechanisms focusing on deterministic and stochastic dynamics, respectively. Section~\ref{sec:indirect} considers the routes by which higher-order effects arise without a direct higher-order interaction. Section~\ref{sec:inference} addresses inference from data, including the identifiability of mechanisms. Section~\ref{sec:synthesis} draws the comparison together and identifies open problems.

%%% ============================================================
\section{Mechanisms, Behaviour and Representation}
\label{sec:reps}
\label{sec:reps:definitions}% mechanistic/behavioural definitions are given in this section
\label{sec:reps:direct}% direct/indirect mechanisms are discussed in this section
%%% ============================================================

Before surveying models, we fix notation and terminology. We consider a system of
$N$ elements, each described by a state variable $s_i$ that may be discrete (for
instance $s_i \in \{0,1\}$ or a finite alphabet) or continuous
($s_i \in \mathbb{R}$). We write $\mathbf{s} = (s_1, \ldots, s_N)$ for the joint
state and $p_N(\mathbf{s}) = p_N(s_1, \ldots, s_N)$ for its joint probability
distribution.

By a \emph{higher-order interaction} we mean a coupling among three or more elements that is not decomposable into a sum of pairwise contributions, an irreducible group effect. Different approaches formalize this criterion in distinct ways, and moving between those formalizations without explicit clarification is a recurrent source of confusion.

\begin{figure}[htbp]
\begin{center}
\includegraphics[width=\textwidth]{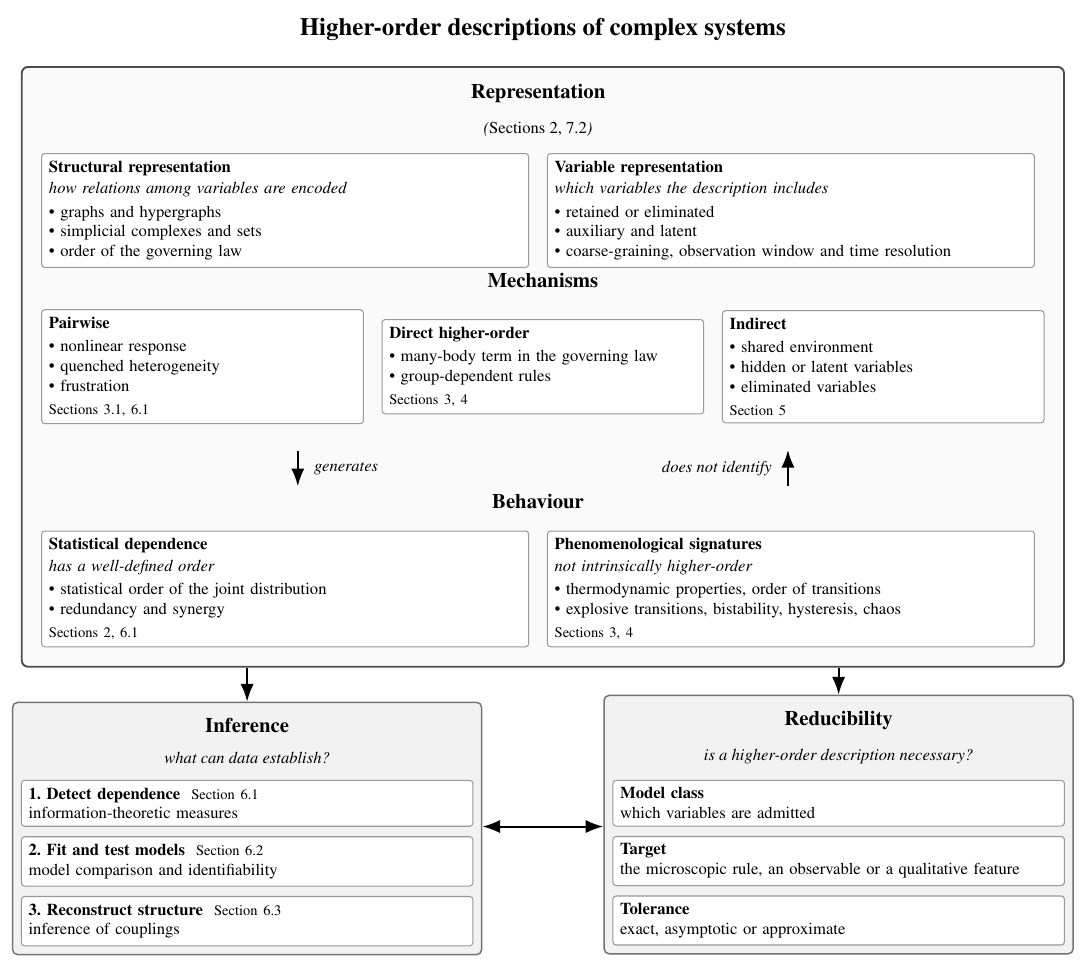}
\end{center}
\caption{Conceptual structure of the review. Higher-order statistical dependence, and the phenomenological
signatures associated with higher-order models, can arise from a pairwise mechanism, a direct many-body mechanism, or
an indirect mechanism acting through a shared environment, a latent variable or
a variable that has been eliminated from the description. Any such mechanism, and the
behaviour it generates, is expressed within a representation, comprising
the structural representation, how the relations among variables are
encoded, and the variable representation, which variables the description
includes. Because distinct mechanisms
can generate overlapping behaviour, the mapping from mechanisms to behaviours is not
one-to-one. Accordingly, an observed dependence or signature does not by itself identify the mechanism
that produced it.
Inference and reducibility address complementary questions: first, what the data
can establish, and second, whether a higher-order description is necessary for a
specified target within a specified model class and tolerance. Labels indicate
the sections in which each element is developed.}
\label{fig:framework}
\end{figure}

The criterion above can be formalized in two ways, according to whether one examines the governing law or the distribution of states. The first focuses on mechanism and the second on behaviour \citep{Rosas2022} (Figure~\ref{fig:framework}). Differentiating them is essential to what follows.

Under the first, \emph{mechanistic} definition, interaction order is determined by the functional form of the governing law, rather than by the number of elements appearing in a particular representation. For a system of $N$ elements with states $\mathbf{s}=(s_1,\ldots,s_N)$ and Hamiltonian $H(\mathbf{s})$, the \emph{mechanistic order} $d$ is the smallest integer for which
\begin{equation}
H(\mathbf{s})=
\sum_{\substack{A\subseteq\{1,\ldots,N\}\\|A|\leq d}}
H_A(\mathbf{s}_A),
\label{eq:hamiltonian}
\end{equation}
where $\mathbf{s}_A$ denotes the states of the elements indexed by $A$. The functions $H_A$ are otherwise unrestricted, and need not individually be irreducible; irreducibility is determined by the minimum order $d$ for which an exact decomposition exists. Equation~\eqref{eq:hamiltonian} must hold for all admissible configurations, with the state variables and any constraints held fixed. Thus, a system has at most pairwise interactions when $d\leq2$, whereas $d\geq3$ indicates irreducible higher-order interactions. Constraints may reduce the order by making an apparently higher-order contribution equivalent to lower-order ones on the admissible state space. For example, for three Ising spins constrained to $s_1+s_2+s_3=1$, the product $s_1s_2s_3$ equals $-1$ on every admissible configuration, so a three-body term reduces to a constant.

For dynamical systems $\dot{s}_i=F_i(\mathbf{s})$, the same principle applies to the decomposition of each $F_i$, counting the affected element $i$ together with those on which each contribution depends. Intrinsic dynamics involving only $s_i$ has order one, an influence of $s_j$ on $i$ has order two, and an irreducible joint influence of $s_j$ and $s_k$ on $i$ has order three. This distinction between nominal interaction structure and irreducible dynamical coupling is also emphasised by Battiston et al.~\citep{Battiston2026}. A closely related definition, stated for the update map of each element and counting only the variables on which a contribution depends, is given by \citet{Rommens2026b}, who show that the order so defined is a property of the map rather than of the structural language used to express it.

Mechanistic order is therefore defined relative to the chosen state variables, admissible configurations and governing law. Changes of variables, the introduction of auxiliary degrees of freedom or coarse-graining may alter it, so irreducibility at one level of description does not necessarily establish an irreducible microscopic mechanism.

Under the second definition, higher-order structure is sought in the
\emph{behaviour} of the system rather than in its governing rule. We distinguish
two modes of characterising behaviour. \emph{Phenomenological signatures} are
collective or macroscopic properties, whether thermodynamic, such as the order of
a phase transition, or dynamical, such as bistability, explosive transitions and
hysteresis (Sections~\ref{sec:deterministic} and~\ref{sec:stochastic}). They carry
no order of their own and are not intrinsically higher-order, even when they arise
in models with higher-order interactions. \emph{Statistical dependence} concerns
the joint distribution $p_N(\mathbf{s})$ of the variables, and in its higher-order
form is an irreducible dependence among three or more of them. For a strictly
positive distribution, the \emph{statistical order} $k$ is the smallest integer
for which
\begin{equation}
\log p_N(\mathbf{s})
=
C+
\sum_{\substack{A\subseteq\{1,\ldots,N\}\\1\leq |A|\leq k}}
\phi_A(\mathbf{s}_A),
\label{eq:statistical_order}
\end{equation}
where the functions $\phi_A$ are unrestricted and the equality holds
for all configurations. Thus, $k\leq 2$ means that the distribution
can be represented using at most pairwise statistical potentials,
whereas $k\geq 3$ requires irreducible higher-order terms.
This factorisation into statistical potentials should not be
confused with factorisation into marginal distributions.
Equivalently, $k$ is the smallest order at which the maximum-entropy
distribution matching all marginals up to that order reproduces $p_N$
exactly~\citep{Amari2001}, which is the form in which the question is posed when models
are fitted to data (Section~\ref{sec:inf:identify}).
The logarithm is natural for both characterizations, since $-\log p_N(\mathbf{s})$ is the information content of configuration $\mathbf{s}$, whose average is the entropy of Section~\ref{sec:inf:measures}, and maximum-entropy distributions are exponentials of sums of the constrained terms.
In equilibrium, where $p_N\propto e^{-\beta H}$, the two orders coincide,
since $\phi_A=-\beta H_A$ and hence $k=d$. Out of equilibrium they need
not, as a generator coupling elements only in pairs may leave a
stationary distribution with $k\geq3$.
The two modes overlap, since many phenomenological signatures are functionals of
$p_N$. Neither statistical dependence nor a phenomenological signature, taken
alone, uniquely identifies the mechanism that produced it.

The distinction between mechanism and behaviour
is loosely analogous to that between statistical mechanics and thermodynamics,
where a macroscopic description stands on its own, autonomously from the
microscopic mechanism that underlies it.

The mechanistic and behavioural definitions are connected but not equivalent. A direct higher-order mechanism, expressed as a genuine
many-body term in the governing law, can induce higher-order statistical dependence and the phenomenological signatures associated with higher-order models.
However, the converse does not hold, since both can be produced without any
direct higher-order interaction. When a lower-order model reproduces such a
dependence or signature without capturing the original higher-order mechanism, we call it
\emph{non-diagnostic}. Such a reduction establishes reducibility with respect to that behaviour alone and not with respect to the mechanism, a distinction made precise below.

Three mechanistic routes to such behaviour are therefore worth distinguishing, and we use them
throughout. A \emph{direct higher-order} mechanism is the many-body term just
described. An \emph{indirect} mechanism couples the units only pairwise, or not
at all, to a shared intermediary, a common stochastic environment or a latent
variable, yet still produces irreducible higher-order statistical dependence among them once
that intermediary is left unobserved (Section~\ref{sec:indirect}). A
\emph{nonlinear but pairwise} mechanism, subject to noise or quenched disorder,
produces phenomenological signatures associated with higher-order models, such as a discontinuous transition,
with no higher-order term in the governing law at
all (Section~\ref{sec:det:sync}).
%%%%%
\emph{Temporal aggregation} stands apart from all three, since no mechanism is involved. Interactions that are genuinely pairwise at each instant can appear as a higher-order group once they are examined at a coarser time resolution. Group interactions in real social data carry their own temporal signatures, with characteristic patterns of formation and dissolution that a coarser time resolution does not necessarily capture~\citep{Cencetti2021}. The time resolution is a feature of the
description rather than of the system, so this route belongs with the
representational choices discussed in
Section~\ref{sec:synthesis:degeneracy}. Where no commitment to any of these
routes is intended, we speak of \emph{higher-order effects}, meaning any statistical
dependence, phenomenological signature or apparent structure usually attributed to
higher-order interactions.
%%%%

In fact, mechanism and behaviour are always described within a \emph{representation}. We
use the term in a broad sense, covering two choices that we name for later use.
The \emph{structural representation} is how the relations among variables are
encoded, and the \emph{variable representation} is which variables are included
in the description at all. The structural representation is the most commonly
studied of the two, and we take it up first.

Both mechanistic and behavioural definitions can be represented as a \emph{hypergraph}, whose hyperedges attach to
groups of three or more nodes~\citep{Battiston2020,Battiston2021,Bianconi2021book}:
a hyperedge may record a mechanistic coupling term or a behavioural relationship
such as a statistical dependency, the combinatorial object being the same in
either case. A \emph{simplicial
complex} is the special case of a hypergraph closed under taking subsets (every
face of a simplex is present), natural in topological settings but restrictive as
a model of interactions. This restriction has led ecological applications such as \citet{Silk2022} to adopt the more general \emph{simplicial set}, which they use to relax the downward-closure requirement while still containing simplicial complexes as a special case. The term is used here in that sense, adapted from algebraic topology, where a simplicial set carries further structure.
A structural representation records a mechanism or a behaviour
rather than constituting a separate kind of higher-order interaction, and the
same mechanism or behaviour can be encoded by more than one of
them~(Section~\ref{sec:synthesis:degeneracy}).

This carries a consequence that is easy to miss. A graph records which elements
interact, not how they interact, and a rule attached to a graph may depend on
the joint state of all the nodes neighbouring a node. Such a rule can be higher-order in
terms of interaction mechanisms even though the structure contains no hyperedge at
all~\citep{Peixoto2026,Battiston2026}. The term \emph{reducible} therefore hides two distinct questions. The first asks whether the same dynamics can be expressed
in a simpler structural language. For example, whether a hypergraph can be replaced by a graph. This is
a question about structural representation and about the informational cost of a modelling
choice~\citep{Peixoto2026,Rommens2026}. The second asks whether the interaction between units in a group can be replaced by one of genuinely lower order. The two questions are distinct, although not independent, as the variable representation now makes clear.

In terms of the variable representation, a
description may retain only some of the degrees of freedom of a system,
introduce auxiliary variables that were not present originally, or fix the scale
and the time resolution at which states are recorded. Each of these choices can
change the interaction order that the description exhibits. Eliminating a
variable that couples to several others leaves an effective many-body term among
those that remain, as the consumer--resource construction of
Section~\ref{sec:det:eco} shows explicitly. Introducing an auxiliary
variable can absorb a many-body term into pairwise couplings, as in the
$p$-spin construction of Section~\ref{sec:stoch:statmech}. The two choices are
therefore coupled, which is what makes the two questions above distinct without
being independent. The variable representation fixes what the structural
representation has to carry, so whether a hypergraph can be replaced by a graph
depends on which variables the description admits rather than on the system
alone. The interaction order attributed to a system is a property of the system
together with both representational choices, a point developed in
Section~\ref{sec:synthesis:degeneracy}. Our discussion concentrates on the
second question while keeping the variable representation it presupposes
explicit.

The \emph{reduction question} can be decomposed into \emph{three subquestions}: (i) which models are admitted as reduced candidates, in particular whether auxiliary variables may be introduced; (ii) what must be reproduced in the reduced version, from the microscopic rule or full dynamics to a macroscopic observable or qualitative feature; and (iii) how exact the agreement between the original and reduced models must be. We use this classification below and return to it in Section~\ref{sec:synthesis}.

Reducibility, as described above, compares models with one another, asking whether a lower-order description can stand in for a higher-order one. A related but distinct question is \emph{identifiability}, which runs from data to models, asking what observations can establish about the model that generated them. The two can return different answers for the same system. This review is concerned principally with reducibility, and takes up identifiability where it bears on that question (Sections~\ref{sec:inf:identify} and~\ref{sec:disc:formal}).

Subquestion (i) is a choice of variable representation, since it fixes which
variables the reduced description may use. Subquestion (ii) connects the mechanism and behaviour definitions given above. To require the full microscopic dynamics is to require the mechanism, whereas to require a set of observables or a qualitative feature is to require only the behaviour. The
weaker the target, the more readily a lower-order model meets it, which is why a
claim resting on a qualitative signature such as the order of a transition is
the easiest to satisfy and the least informative.

%%% ============================================================
\section{Deterministic Models}
\label{sec:deterministic}
%%% ============================================================

Higher-order interactions as generalizations of pairwise coupling have been studied extensively using deterministic models defined by coupled ordinary differential equations or maps. We focus on three representative cases: synchronization, ecology and evolution.

\subsection{Synchronization}
\label{sec:det:sync}

The Kuramoto model is the main paradigm for
synchronization~\citep{Acebron2005}.
In its pairwise form, $N$ phase oscillators evolve as
\begin{equation}
  \dot{\theta}_l \;=\; \omega_l
    + \frac{K}{N}\sum_{j=1}^{N}\sin(\theta_j - \theta_l),
  \label{eq:kuramoto}
\end{equation}
with natural frequencies $\omega_l$ drawn from a distribution $g(\omega)$.
 The
degree of synchronization is measured by the complex order parameter, or
mean field,
\begin{equation}
  Z \;=\; R\,e^{i\psi} \;=\; \frac{1}{N}\sum_{j=1}^{N} e^{i\theta_j},
  \label{eq:korder}
\end{equation}
whose modulus $R$ vanishes when the phases are incoherent and approaches unity
for complete locking. In terms of $Z$, equation~\eqref{eq:kuramoto} becomes
$\dot{\theta}_l = \omega_l + K\,\mathrm{Im}\!\left[Z\,e^{-i\theta_l}\right]$,
so that each oscillator is coupled to the rest only through the mean field. For a unimodal and symmetric
$g(\omega)$ satisfying $g''(0)<0$, the incoherent state $R=0$ is stable up to a critical coupling
$K_c = 2/[\pi g(0)]$, beyond which a synchronized branch grows continuously as
$R \sim (K-K_c)^{1/2}$. The bifurcation of the amplitude $R$ is supercritical, so
that the onset of synchronization is continuous. In contrast, the bifurcation becomes subcritical when
$g''(0)>0$~\citep{Acebron2005}.

Higher-order generalizations of the Kuramoto model~\citep{SkardalArenas2020,Bick2023}
add triadic and tetradic terms, for instance
\begin{equation}
  \dot{\theta}_l \;=\; \omega_l
    + \frac{K_1}{N}\sum_{j}\sin(\theta_j - \theta_l)
    + \frac{K_2}{N^2}\sum_{j,k}\sin(2\theta_j - \theta_k - \theta_l)
    + \frac{K_3}{N^3}\sum_{j,k,n}\sin(\theta_j + \theta_k - \theta_n - \theta_l).
  \label{eq:hokuramoto}
\end{equation}
This system reduces to~\eqref{eq:kuramoto} with $K = K_1$ when $K_2 = K_3 = 0$.
In its general form, the model robustly displays \emph{explosive synchronization} even when $g''(0)<0$, namely a discontinuous,
hysteretic jump in $R$ as the coupling is varied, together with regions of
bistability between the incoherent and synchronized states. This is linked to a change in the bifurcation structure: for sufficiently large $K_2+K_3$ the bifurcation becomes subcritical and a saddle-node creates a bistable interval~\citep{SkardalArenas2020}.

The appearance of an explosive transition upon activating higher-order
interactions in~\eqref{eq:hokuramoto} might suggest that they are necessary for
such phenomenology in systems of oscillators. However, higher-order interactions are not essential to observe this macroscopic behaviour.

Indeed, explosive synchronization arises in models whose
interaction rule is strictly pairwise. This occurs, for example, in models where the symmetry between oscillators
is broken so that the frequency of a node is tied to its effective coupling~\citep{GomezGardenes2011,Zhang2013}.
In these models, the coupling
term retains the pairwise form $\sum_j K_{lj}\,\sin(\theta_j-\theta_l)$ with coefficients
$K_{lj}$ that are fixed in advance and do not depend on the state of the system, so the explosive transition is produced by heterogeneity quenched into the node parameters.
What these pairwise models
reproduce is a qualitative feature of the collective behaviour
of~\eqref{eq:hokuramoto}, the order of the transition and its hysteresis, rather
than its dynamics or any particular observable of it. This is the
weakest of the options listed under subquestion (ii) of
Section~\ref{sec:reps}, which makes explosive synchronization non-diagnostic in
the sense given there.

A stronger requirement under the reducibility subquestion (ii) of Section~\ref{sec:reps} is worth exploring here for the
all-to-all model. \citet{Llabres2026} call a
higher-order model \emph{macroscopically} reducible when a lower-order model
obeys the same evolution equation for a macroscopic observable. They demonstrate such reducibility for the linear voter model on hypergraphs, whose density of
active links evolves exactly as on the projected graph. The mean-field
reduction of the all-to-all Kuramoto model allows the same comparison to be
made in terms of the macroscopic quantity $R$.
Indeed, the Ott--Antonsen ansatz~\citep{OttAntonsen2008} applies to
populations whose velocity field is a \emph{first harmonic} in the phase of each oscillator, $\dot{\theta}_l = \omega_l + \mathrm{Im}[A e^{-i\theta_l}]$ with $A$ a
function of global quantities alone. The pairwise
model~\eqref{eq:kuramoto} is the case $A = KZ$. The couplings
in~\eqref{eq:hokuramoto} are constructed to satisfy this condition, the three-
and four-body sums collapsing into products of the order parameters
$Z_m = N^{-1}\sum_j e^{im\theta_j}$. For a Lorentzian $g(\omega)$, which further
slaves the higher harmonics to the first through $Z_m = Z_1^{m}$,
\citet{SkardalArenas2020} thereby reduce~\eqref{eq:hokuramoto}
to a Kuramoto model whose coupling strength, $K_{\mathrm{eff}}(R) = K_1 + (K_2+K_3)\,R^{2}$, is modulated by the global order
parameter. Specifically, setting $\omega_0=0$, the modulus $R$ obeys
\begin{equation}
  \dot{R} \;=\; - R \Delta\;+\; \tfrac{1}{2}\,K_{\mathrm{eff}}(R)\,R\left(1-R^{2}\right),
  \label{eq:amplitude}
\end{equation}
with $\Delta$ the half-width of the Lorentzian. The pairwise model~\eqref{eq:kuramoto} obeys the same equation with the constant value $K_{\mathrm{eff}} = K$, so the two models differ only through the $R$ dependence of $K_{\mathrm{eff}}$.

This shows that the many-body dependence is confined to a single collective
variable, the triadic and tetradic couplings acting only through $R$ and only in
the combination $K_2+K_3$. What they contribute macroscopically is a coupling
that grows with the coherence of the population, and it is this feedback between
order and coupling strength that turns the transition discontinuous.

The reduction leading to Eq.~\eqref{eq:amplitude} does not, however, remove the
many-body character of the interaction. Since $R$ depends on the phases of every
oscillator, a model with coupling $K_{\mathrm{eff}}(R)$ is not pairwise by the
mechanistic criterion of Section~\ref{sec:reps}. The higher-order Kuramoto model is
therefore not reducible to the pairwise Kuramoto model in terms of the
macroscopic observable $R$ in the sense of \citet{Llabres2026}, since no constant
value of $K_{\mathrm{eff}}$ reproduces $K_1 + (K_2+K_3)R^{2}$. That $R$
dependence also has important consequences for the dynamics, since the
supercritical bifurcation exhibited by the pairwise model becomes subcritical for
$K_2+K_3 > K_1$. This does not conflict with the pairwise models discussed above,
which reproduce the explosive behaviour but are not of the form of
Eq.~\eqref{eq:kuramoto}.

Not every route to this feedback is itself pairwise. The quenched heterogeneity
of the pairwise models above supplies it by other means, whereas an adaptive
coupling set by a local order parameter~\citep{Zhang2015} supplies it through a
rule that depends on the joint state of a neighbourhood, and so fails the same
mechanistic criterion.

The preceding reductions assume all-to-all coupling. In addition, the Ott--Antonsen collapse further requires sinusoidal interaction, a specific frequency distribution and $N\to\infty$. For heterogeneous frequencies, the reduced manifold is invariant and attracting, so the reduction captures the long-time dynamics of the order parameter rather than arbitrary transients~\citep{OttAntonsen2009}. However, explosive transitions persist on structured networks, where the all-to-all condition is relaxed~\citep{MillanTorresBianconi2020}, and linear stability can still be analysed by combining interaction-order Laplacians~\citep{LucasBattiston2020,Gambuzza2021}. With an unbiased rescaling of the coupling strengths across orders, a purely pairwise network can be tuned to match the linear stability of a given higher-order network exactly~\citep{Ren2023}. Apparent order-dependent effects on synchronization stability can therefore be an artefact of how couplings were normalized, rather than evidence of a genuine higher-order mechanism.

The reducibility question remains open where mean-field closure fails or multistability, heteroclinic cycles and chaos occur~\citep{Bick2023,MillanTorresBianconi2020,LucasBattiston2020}. The purely three-body model of \citet{SkardalArenas2019}, for example, admits only a partial reduction, since only the symmetric part of the oscillator density collapses onto an Ott--Antonsen manifold and the remainder has to be closed by a separate self-consistency argument. The asymmetry left outside the reduction is what indexes the extensive family of multistable states the model displays. Linear stability is nevertheless settled by the Laplacian mapping, so any irreducibility must arise in the nonlinear regime.

Reducibility depends on whether group contributions are additive or genuinely non-additive. Nonlinear rules can relocate rather than remove many-body dependence, for example through configuration-dependent couplings~\citep{Neuhauser2020,Sahasrabuddhe2021,Llabres2026}.

Neither the microscopic nor the macroscopic reduction succeeds in the higher-order Kuramoto
case. However, this is not the rule in general, and both succeed in other settings. Linear consensus dynamics on a hypergraph maps exactly onto pairwise
dynamics on a weighted projected
network~\citep{Neuhauser2020,Sahasrabuddhe2021,Llabres2026}, and for the linear
voter model the density of active links then obeys the same evolution equation
as the standard voter model on the projected network with \emph{unweighted}
links~\citep{Llabres2026}, so the model reduces at both micro and macroscopic levels.

\subsection{Ecology and Evolution}
\label{sec:det:eco}

Population ecology and evolutionary dynamics are the examples we consider next within a deterministic
setting. Higher-order interactions have a
particularly transparent appearance through the generalized Lotka--Volterra model with higher-order terms,
\begin{equation}
  \dot{x}_i \;=\; x_i\Bigl( r_i + \sum_j \alpha_{ij} x_j
      + \sum_{j,k} \beta_{ijk}\, x_j x_k + \cdots \Bigr),
  \label{eq:glv}
\end{equation}
that governs the abundance $x_i \geq 0$ of type $i = 1,\ldots,N$, a species in
ecology or the subpopulation using strategy $i$ in evolutionary dynamics. Here, $r_i$ is the intrinsic
growth rate of type $i$, while $\alpha_{ij}$ quantifies the effect of type $j$
on the growth of type $i$. The higher-order terms form a hierarchy in which $\beta_{ijk}$
encodes a genuine three-body effect, type $k$ modifying the interaction between the types $i$ and $j$.

A separate choice, made in both ecology and evolution, is whether to work with absolute or
normalized abundances. Writing $p_i = x_i/\sum_j x_j$, so that $\sum_i p_i = 1$, places the state on the probability simplex and relates the Lotka--Volterra dynamics to replicator dynamics~\citep{HofbauerSigmund1998}. Among the ecological models below, \citet{Grilli2017} and \citet{Bairey2016} impose
$\sum_i x_i = 1$ whereas \citet{Gibbs2022} work with unconstrained abundances.
 Normalization changes how
higher-order terms scale with system size and can itself induce statistical
dependence among variables, so it is a choice of variable representation in the
sense of Section~\ref{sec:reps} rather than merely a change of units.

\subsubsection{Ecology: Impact of higher-order interaction on species stability and reducibility}

The ecological interest in higher-order interactions is driven by their impact on species coexistence. The natural baseline framework in this context is the pairwise theory of community stability due to
\citet{May1972}. A random community of $N$ species with connectance $C$ and interaction strength $\sigma$ is almost surely stable only if $\sigma\sqrt{NC}<1$. Since the left-hand side grows with both $N$ and $C$, increasing diversity or
connectance \emph{destabilizes} the community. Biodiversity then appears paradoxical since species-rich communities should not persist~\citep{May1972,May2001}. Higher-order interactions were proposed as a way out, and their consequences were modelled before much empirical evidence for them was available.

Both \citet{Grilli2017} and \citet{Bairey2016} report that higher-order interactions favour
coexistence: the former in competitive network models generalizing
rock--paper--scissors dynamics, the latter by inverting May's
diversity--stability relationship, so that the community becomes sensitive to
the removal of species rather than to their addition.

Stabilization is not the general rule, however, and \citet{Levine2017} caution
that higher-order mechanisms need not stabilize coexistence and may destabilize
it. Following the spirit in which May treated $\alpha_{ij}$, \citet{Gibbs2022} proposed to draw the $\beta_{ijk}$ at random, with controlled mean and variance across species. The result is a concrete instance of that caution. Asking how many of the
abundances $x_i$ in~\eqref{eq:glv} can coexist, they find that greater
variability in the $\beta_{ijk}$ reduces coexistence and that species-rich
communities lose species more readily than species-poor ones, so May's
diversity--stability relation is extended to the higher-order case rather than
inverted.

\citet{Gibbs2022} trace the opposing conclusions to normalization. With abundances summing to one, as in \citet{Bairey2016}, each $x_i=O(1/N)$. Each additional abundance in a product then lowers the weight of a term by a factor of order $N$, so higher-order interactions weaken with diversity. With unconstrained abundances, as in~\eqref{eq:glv}, and with the standard deviations of $\alpha_{ij}$ and $\beta_{ijk}$ scaled as $N^{-1/2}$ and $N^{-1}$, as assumed by \citet{Gibbs2022}, higher-order variability affects coexistence as pairwise variability does. Normalization therefore changes the inferred role of higher-order interactions.

Whether higher-order interactions enhance or diminish coexistence also depends on how they are specified, including their functional form and their correlation with the pairwise coefficients~\citep{Gibbs2022,Gibbs2024}.

Empirical tests were long rare enough to leave the importance of these
mechanisms in nature unknown~\citep{Levine2017}, although they have begun to
appear. One in natural plant communities found that variation in seed production
was explained substantially better once higher-order interactions were
included~\citep{Mayfield2017}. Subsequent mesocosm and microcosm experiments
have directly measured higher-order effects on species persistence in
multitrophic plant--pollinator and aquatic protist
communities~\citep{BucheGodoy2024,Shen2023}. At a larger scale, higher-order terms improve predictions of tree growth and survival across forest plots~\citep{Li2026}.

\citet{Levine2017} also distinguish two kinds of dynamics that arise only among three or more
competitors. In \emph{interaction chains}, one species affects another through changes in the density of a third, and the interactions remain pairwise. \emph{Higher-order interactions} occur when the per
capita effect of one competitor on another depends on the density of a third,
which is the mechanistic criterion of Section~\ref{sec:reps} stated in
ecological terms.

Under certain conditions, higher-order interactions between species arise from a consumer--resource mechanism, in which species compete indirectly by sharing resources~\citep{cui2024leshoucheslecturescommunity}. To show this, consider $N$ consumers of abundance $x_i$ feeding on $M$ resources of abundance
$R_\mu$, with no direct interaction of any order among the consumers,
\begin{align}
  \dot{x}_i &= x_i\Bigl(\sum_\mu c_{i\mu} R_\mu - m_i\Bigr),
  \label{eq:crm_consumer}\\
  \varepsilon\,\dot{R}_\mu &= s_\mu - \delta_\mu R_\mu
     - R_\mu \sum_j c_{j\mu} x_j.
  \label{eq:crm_resource}
\end{align}
Here, $c_{i\mu}$ is the rate at which consumer $i$ uses resource $\mu$, $m_i$ is
its mortality rate, $s_\mu$ and $\delta_\mu$ are the supply and decay rates of
resource $\mu$, and $\varepsilon$ sets the timescale of the resource dynamics
relative to that of the consumers.

If the resources equilibrate rapidly, $\varepsilon \to 0$, their degrees of
freedom can be adiabatically eliminated by setting $\dot{R}_\mu = 0$, which
gives
\begin{equation}
  R_\mu^{*}(\mathbf{x}) \;=\; \frac{s_\mu}{\delta_\mu + \sum_j c_{j\mu} x_j}.
  \label{eq:crm_qss}
\end{equation}
The quasi-stationary level of each resource depends on the abundances of
\emph{all} consumers exploiting it, and does so nonlinearly. Substituting
\eqref{eq:crm_qss} into \eqref{eq:crm_consumer} closes the dynamics on the
consumers alone, and expanding in powers of the abundances, which is legitimate
when $\sum_j c_{j\mu} x_j \ll \delta_\mu$, reproduces \eqref{eq:glv} with
\begin{equation}
  r_i = \sum_\mu \frac{s_\mu c_{i\mu}}{\delta_\mu} - m_i, \qquad
  \alpha_{ij} = -\sum_\mu \frac{s_\mu}{\delta_\mu^{2}}\,c_{i\mu}c_{j\mu}, \qquad
  \beta_{ijk} = \sum_\mu \frac{s_\mu}{\delta_\mu^{3}}\,c_{i\mu}c_{j\mu}c_{k\mu}.
  \label{eq:crm_coeffs}
\end{equation}

The pairwise coefficient $\alpha_{ij}$ is the classical overlap of the
resource-utilization profiles of species $i$ and $j$, and is negative, so that
consumers sharing resources compete. The higher-order coefficient $\beta_{ijk}$
is the corresponding triple overlap and is positive, so that the leading
higher-order correction weakens competition as the community becomes denser.
Neither coefficient represents a mechanism acting between the species concerned:
both are generated by eliminating the resources, which are variables absent
from~\eqref{eq:glv}.
This is the indirect mechanism of Sections~\ref{sec:reps:direct}
and~\ref{sec:indirect} in concrete form.

The higher-order Lotka--Volterra model~\eqref{eq:glv} is therefore irreducible
to a pairwise model on the consumer abundances $\mathbf{x}$ and reducible to one
on the enlarged set $(\mathbf{x},\mathbf{R})$, in which every term
of~\eqref{eq:crm_consumer} and~\eqref{eq:crm_resource} couples at most two
dynamical variables. The answer turns on whether the eliminated variables may be reintroduced,
which is a choice of variable representation rather than a fact about the
community.

Resource elimination generates interactions at every order, so Eq.~\eqref{eq:glv} is a truncation of Eq.~\eqref{eq:crm_qss}, and which order suffices depends on the accuracy required, i.e., it is related to subquestion (iii) of Section~\ref{sec:reps}.

In ecology, $\beta_{ijk}$ is frequently defined as the
measured deviation of community dynamics from the additive pairwise prediction,
rather than as an independently established mechanism. Indirect competition
produces such a deviation, but direct group interaction is also possible \citep{Levine2017}. Whether a fitted coefficient reflects a
genuine higher-order mechanism is therefore an identifiability question
(Section~\ref{sec:inference}), and two reductions compete for it. The
shared-resource route is exact but enlarges the variable set (subquestion (i) of Section~\ref{sec:reps}). An effective pairwise model with
nonlinear functional responses instead keeps the original variables and
reproduces the abundance time series only to within a tolerance (subquestions (ii) and (iii)). In fact, higher-order Lotka--Volterra models can be fitted accurately with pairwise descriptions, so that higher-order mechanisms need not be identifiable from abundance time series when flexible pairwise models reproduce the same observed dynamics~\citep{CallejaSolanas2026}. Identifying the best route to describe the system is an experimental rather than a purely observational task.

\subsubsection{Evolution of cooperation with higher-order interactions}

The replicator form is also the setting of evolutionary game theory, where
higher-order interactions enter as multiplayer games. A $d$-player game, of
which the public-goods game is the standard example, assigns payoffs to the
composition of a group and is higher-order by construction, supporting internal
equilibria and coexistence patterns unreachable with two-player
payoffs~\citep{Gokhale2010}. On higher-order structured populations,
\citet{Guo2021} report a transition from dominant defection to dominant
cooperation on simplicial complexes as group interactions become more prevalent,
together with the coexistence of dominant and non-dominant strategies.
\citet{Civilini2024} report an explosive transition to cooperation on hypergraphs, with
a bistable regime requiring an initial critical mass of cooperators.

Neither study is framed as a reducibility test. \citet{Guo2021} state that the coexistence they observe cannot be reproduced by pairwise games on any network of contacts, but compare with a particular pairwise baseline rather than with a class of pairwise models. In fact, spatial structure is known to promote coexistence in pairwise games~\citep{NowakMay1992}. The explosive transition of \citet{Civilini2024} is a phenomenological signature and, as in Section~\ref{sec:det:sync}, non-diagnostic, since a discontinuous onset with a bistable regime is generic to cooperative nonlinear dynamics.

%%% ============================================================
\section{Stochastic-Process Models}
\label{sec:stochastic}
%%% ============================================================

We turn to models in which the state evolves stochastically. Higher-order
interactions have been studied in a wide range of such settings, among them
neural population activity~\citep{Schneidman2006,Tkacik2014}, opinion and
consensus dynamics on hypergraphs~\citep{Noonan2021,Llabres2026}, and random
walks and diffusion on higher-order structures~\citep{Carletti2020}. We do not
attempt to cover all of them. Instead, we develop two cases in detail, the equilibrium
statistical mechanics of disordered systems and contagion and percolation
processes on networks and hypergraphs. These examples allow the questions
of Section~\ref{sec:reps} to be posed
concretely: in the first, the interaction order is an explicit control parameter
of the model, whereas in the second, the same observations are accounted for by
different mechanisms, some carried by hyperedges and others by higher-order rules on
an ordinary graph.

\subsection{Statistical Mechanics: The \texorpdfstring{$p$}{p}-Spin Glass}

\label{sec:stoch:statmech}

In the statistical mechanics of disordered systems the effect of interaction
order can be established precisely in terms of the $p$-spin model defined by the Hamiltonian~\citep{Derrida1980,GrossMezard1984}
\begin{equation}
  H(\mathbf{s}) \;=\; -\!\!\sum_{i_1 < \cdots < i_p}\! J_{i_1\cdots i_p}\,
      s_{i_1}\cdots s_{i_p}.
  \label{eq:pspin}
\end{equation}
Here, $s_i \in \{-1,+1\}$. The couplings $J_{i_1\cdots i_p}$ are quenched Gaussian random variables,
fixed for each realization of the disorder and determining the Hamiltonian. In the
terms of Section~\ref{sec:reps:definitions}, they represent the interaction
mechanism. Each term in~\eqref{eq:pspin} couples exactly $p$ spins. The spins
$s_i$ evolve stochastically, through single-spin Glauber or Metropolis updates
satisfying detailed balance at temperature $T$, so that the stationary
distribution is the Gibbs measure
\begin{equation}
p_N(\mathbf{s}) \propto e^{-H(\mathbf{s})/k_B T}.
\label{eq:pN_pspin}
\end{equation}

For $p\geq3$ the thermodynamics differs qualitatively from the pairwise Sherrington--Kirkpatrick model. In the standard mean-field treatment, the $p=2$ case exhibits full replica-symmetry breaking, whereas $p\geq3$ exhibits one-step RSB, a discontinuous Edwards--Anderson order parameter and exponentially many well-separated pure states~\citep{GrossMezard1984,Derrida1980,CastellaniCavagna2005,MezardMontanari2009}. With Ising spins, the one-step solution describes the glass phase just below the transition but becomes unstable at lower temperature towards full replica-symmetry breaking, the Gardner transition~\citep{Gardner1985,Gross1985}. There are distinct dynamical and thermodynamic transitions, the hallmark of the random first-order transition (RFOT) scenario. Beyond $p=3$ the change is quantitative rather than qualitative. The covariance of the energies of two configurations depends on their overlap $q$ as $q^{p}$, so increasing $p$ leaves the energies of distinct configurations progressively less correlated, until as $p\to\infty$ they become independent and the model reduces to the random-energy model~\citep{Derrida1980}.

Concerning reducibility, let $\mathcal{M}_1$ be the set of pairwise Hamiltonians over the $N$ spins of the original $p$-spin model. No such Hamiltonian reproduces the $p$-body interaction structure of the $p\geq3$ model, so the model is irreducible within $\mathcal{M}_1$. Its phenomenological signatures, however, is non-diagnostic in the sense of Section~\ref{sec:reps}. One-step RSB and the RFOT scenario also arise from pairwise couplings, as in the Super-Potts model, whose $n$-state variables interact in pairs yet display one-step RSB and RFOT behaviour for large enough $n$~\citep{AngeliniBiroli2014}. Reducibility is therefore relative to the target. Taking the $p$-body interaction structure as the target leaves the model irreducible within $\mathcal{M}_1$, whereas qualitative features such as one-step RSB or RFOT behaviour do not by themselves require a higher-order interaction.

A different reduction is possible for pairwise models $\mathcal{M}_2$ over an enlarged variable set (i.e. the original $N$ spins plus auxiliary variables). In a factor-graph representation of the $p$-spin model, each $p$-spin term of the Gibbs weight~\eqref{eq:pN_pspin} is a single factor node of degree $p$, connected to the $p$ spin variables it couples~\citep{MezardMontanari2009}. Such a factor is a function of all $p$ spins simultaneously,
\[
\psi(s_{i_1},\ldots,s_{i_p})
=
\exp\!\left(
\beta J_{i_1\cdots i_p}
s_{i_1}\cdots s_{i_p}
\right),
\qquad
\beta=(k_B T)^{-1},
\]
and encodes a genuine $p$-body interaction, since it does not factorize into a product of pairwise terms.

Now replace that factor node by an auxiliary variable $z=(z_1,\ldots,z_p)$ with
$z_k\in\{-1,+1\}$, whose $2^p$ states correspond one-to-one with the
configurations of the $p$ spins attached to the factor. Give it the
one-variable weight $h(z)=\exp(\beta J_{i_1\cdots i_p}z_1\cdots z_p)$, and
connect it separately to each spin through the pairwise compatibility function
$\varphi_k(s_{i_k},z)$, equal to one when $s_{i_k}=z_k$ and zero otherwise. The
original factor is then recovered exactly by marginalizing $z$:
\begin{equation}
\sum_{z} h(z)
\prod_{k=1}^{p}\varphi_k(s_{i_k},z)
=
\psi(s_{i_1},\ldots,s_{i_p}).
\label{eq:auxmarg}
\end{equation}
Indeed, the product of compatibility functions is non-zero for exactly one
state of $z$, namely $z=(s_{i_1},\ldots,s_{i_p})$, so
that~\eqref{eq:auxmarg} simply evaluates $h(z)$ at the observed spin
configuration. This is an identity between equilibrium weights: it reproduces
the partition function, the free energy and the static correlations of the
observed spins, but says nothing about the stochastic dynamics, which the
Hamiltonian alone does not fix. Every interaction in the enlarged graph thus involves at most two variables. Read in the opposite direction, this
shows that integrating out a hidden variable coupled to several observed
variables can generate an effective higher-order interaction among those
variables, as in Boltzmann machines with hidden units~\citep{Ackley1985}. Since their pairwise parameters grow only quadratically with the number of units, matching an arbitrary distribution requires exponentially many hidden units~\citep{Ackley1985,LeRouxBengio2008}.

The representation is pairwise only because $z$ counts as a \emph{single} variable with $2^p$ states, so that $-J_{i_1\cdots i_p}z_1\cdots z_p$ is a one-variable energy. Were its components counted as separate binary variables, the $p$-body term would simply have moved to the auxiliary ones. The reduction therefore depends on what is an admissible variable in $\mathcal{M}_2$ (Section~\ref{sec:reps}). Its cost is informational, namely the $p$ bits carried by the $2^p$ states of $z$, and is quantified in general terms by \citet{Rommens2026}. It does not imply that an arbitrary $p$-spin model can be represented by an ordinary pairwise Ising model of comparable size.

Random $k$-satisfiability is a non-dynamical analogue~\citep{MezardMontanari2009}. Each clause of a $k$-SAT formula is an irreducible $k$-body term of the energy that counts violated clauses, and the random ensemble undergoes clustering transitions closely related to those of the $p$-spin model. Any $k$-SAT instance can be rewritten as an equisatisfiable $3$-SAT instance using auxiliary Boolean variables, whereas $3$-SAT cannot be reduced to $2$-SAT unless $\mathrm{P}=\mathrm{NP}$, so the irreducibility of the order is certified by complexity theory.

\subsection{Contagion and Percolation}
\label{sec:stoch:contagion}

Higher-order contagion models differ in where their interacting groups come from. Some use hyperedges or simplices as additional structure; others induce groups from an ordinary graph, such as the neighbourhood of a node. Both can implement mechanistically higher-order transmission, so the architecture itself is not the essential distinction.

\subsubsection{Contagion on higher-order architectures}
The simplicial susceptible--infected--susceptible (SIS) model of
\citet{Iacopini2019} augments pairwise transmission with a higher-order channel. In addition to ordinary pairwise infection at rate $\beta$, an infected pair jointly infects a third node of a filled triangle, a 2-simplex, at rate $\beta_\triangle$.
In the well-mixed mean-field limit, the infected fraction $\rho$ obeys an
equation of the form
\begin{equation}
  \dot{\rho} \;=\; -\mu\rho
    + \beta \langle k\rangle\,\rho(1-\rho)
    + \beta_\triangle \langle k_\triangle\rangle\,\rho^{2}(1-\rho),
  \label{eq:simplicialsis}
\end{equation}
where $\mu$ is the recovery rate; $\langle k\rangle$ and $\langle k_\triangle\rangle$
count pairwise and triangular neighbourhoods, respectively.
The cubic term produces a saddle--node bifurcation. This leads to a discontinuous
transition and a bistable region in which an endemic state persists below the
pairwise epidemic threshold~\citep{Iacopini2019,LandryRestrepo2020}.
The construction extends to general hypergraphs and SIR dynamics~\citep{LandryRestrepo2020,deArruda2020,StOnge2021}; in percolation, group activation similarly yields discontinuous giant-component emergence~\citep{SunBianconi2021,Bianconi2021book}. Bistability, discontinuous thresholds and hysteresis recur across these architectures.

The same signatures arise when groups are induced by an ordinary graph, as reported in works predating those using higher-order architectures. In synergistic contagion, transmission depends on the number of infected neighbours, so the neighbourhood jointly modulates infection even without hyperedges~\citep{PerezRecheLudlam2011}. This mechanism produces explosive transitions on networks and rich bifurcation structure on heterogeneous networks~\citep{GomezGardenes2016,Taraskin2019,PerezReche2025}. Threshold and complex-contagion models express the same principle~\citep{Granovetter1978,Watts2002,DoddsWatts2004,CentolaEguiluzMacy2007}. Thus bistability, hysteresis and explosive transitions do not require higher-order architecture: they require a higher-order mechanism, which a graph-based rule can provide.

\subsubsection{Reducibility of contagion models}
In terms of model reducibility, the three targets of comparison listed under subquestion (ii) of
Section~\ref{sec:reps} give three different answers here.

At the level of the \emph{microscopic rule}, the question has an exact answer. A
susceptible--infected process on a hypergraph coincides with one on the
projected graph precisely when the group transmission rate is linear in the
number of infected members, that is when each infective transmits
independently~\citep{Xie2026}. Nonlinearity in that count is what distinguishes
the two, and it is supplied equally by filled triangles and by a synergistic
rule on a graph. Where the rate is nonlinear the compensating pairwise rate must
track the instantaneous number of infected neighbours, so the many-body
dependence is relocated rather than removed, as in Section~\ref{sec:det:sync}.

At the level of a \emph{chosen observable}, here the mean-field density $\rho$, different microscopic architectures can become indistinguishable. In~\eqref{eq:simplicialsis}, the cubic term $\rho^2(1-\rho)$ arises from the nonlinear, group-dependent transmission rule associated with the higher-order network structure. At the level of the mean-field density, however, the same type of nonlinear reinforcement can be generated by synergistic or threshold contagion on a pairwise graph. The resulting equation for $\rho$ therefore does not reveal whether its nonlinearity originated from the higher-order network architecture or from a pairwise network with a nonlinear transmission rule.

At the level of a \emph{qualitative feature}, such as the discontinuous threshold, bistability and hysteresis described above, discrimination is weaker still. Coarse-grained simplicial contagion can have the same form as synergistic or threshold contagion on a graph, higher-order simplices can become irrelevant under renormalization~\citep{Meloni2026}, and nonlinear rate equations can produce multistability and complex bifurcations under different microscopic mechanisms~\citep{PerezReche2025}. These qualitative signatures therefore do not by themselves identify either the microscopic mechanism or the structural representation used to describe it. In particular, a transmission rule that depends nonlinearly on the number or joint state of infected neighbours constitutes a higher-order mechanism irrespective of whether the underlying system is represented by a graph, hypergraph or simplicial complex. What may remain unresolved is which interaction architecture or representation is appropriate for the system. Settling this question requires inference from sufficiently resolved microscopic data (Section~\ref{sec:inference}), rather than macroscopic phenomenology alone.

In another graph-based model, triadic percolation, a node activates or suppresses the link between two others, so the group controls edge \emph{existence} rather than transmission. The resulting percolation process has a time-dependent giant component with period doubling and a route to chaos~\citep{SunRadicchi2023,Peixoto2026}.

The reducibility question remains open for triadic percolation. Fixed-edge pairwise percolation cannot produce a time-dependent giant component, but period doubling and chaos are generic nonlinear phenomena, produced by many mechanisms besides a genuinely higher-order one, and therefore non-diagnostic. Moreover, the reducibility comparisons made elsewhere in this review concern local rules, whereas the triadic update depends on the global giant component, so a definitive lower-order comparison is still lacking.

%%% ============================================================
% REVISED TITLE (R2.1). Was: Higher-Order Behaviour Without Higher-Order Interactions.
% Colour markup cannot be used here because the class uppercases section titles.
\section{Higher-Order Effects Without Direct Higher-Order Interactions}
\label{sec:indirect}
%%% ============================================================

Sections~\ref{sec:deterministic} and~\ref{sec:stochastic} dealt with direct
mechanisms, a many-body term appearing explicitly in the governing law. This
section takes cases in which higher-order effects arise without such a
term. One route has already appeared, the nonlinear but pairwise mechanism of
Section~\ref{sec:det:sync}, where a phenomenological signature,
a discontinuous transition, follows from
heterogeneity or feedback rather than from any group coupling~\citep{GomezGardenes2011,Zhang2013}. The following two routes involve indirect mechanisms that do not belong specifically to either the deterministic or stochastic setting, and are therefore considered separately.

The first is a \emph{shared environment} or common latent driver.
When otherwise independent variables are coupled to a common fluctuating
influence, marginalizing over that influence induces statistical dependence
among them, and, in general, higher-order dependence. A recent minimal
model~\citep{Env2026} makes this precise for variables coupled to a shared
stochastic environment with no direct interactions of any order. The
dependencies it produces are quantified with the information-theoretic measures
of Section~\ref{sec:inf:measures}, where the model is taken up in detail.

The second is \emph{elimination of variables}, the same operation stated in
general terms. As Section~\ref{sec:stoch:statmech} showed, integrating out an
auxiliary node that couples pairwise to several observed variables leaves an
effective higher-order factor among the observables, so apparent higher-order
structure can be the shadow of pairwise coupling to something unmeasured.
Marginalization is one way of eliminating a variable and adiabatic approximation
another, as in the consumer--resource reduction of Section~\ref{sec:det:eco},
where the variable removed is real and physical rather than hidden.
Adiabatic elimination is, however, a limiting case. In general, eliminating
variables yields equations for the retained ones that contain a memory term and a
fluctuating force~\citep{Zwanzig1961,Mori1965,Chorin2000}. Only when the eliminated
variables are fast, as for $\varepsilon\to0$ in Section~\ref{sec:det:eco}, does the
memory collapse into an instantaneous effective interaction, so the interaction order
of an effective description presupposes this time-local closure.

Another source of apparent higher-order structure is \emph{time aggregation}, introduced in
Section~\ref{sec:reps}. It differs from the previous two in that no
mechanism is involved.
%In spite of that, the associated phenomenology can be indistinguishable from that of a direct mechanism, a point developed in Sections~\ref{sec:inference} and~\ref{sec:synthesis}.

The common lesson is that neither higher-order dependence, by whatever
measure, nor the phenomenological signatures often found in models with
higher-order interactions establishes a higher-order mechanism. Discriminating direct
higher-order interactions from these other routes requires leverage that
correlational measures and macroscopic signatures do not provide: interventions that break a putative
shared driver, observation of the candidate hidden variables, or temporal
resolution fine enough to separate simultaneous from sequential coupling. This
returns the question from observed behaviour to the causal and dynamical considerations
of Section~\ref{sec:synthesis}.

%%% ============================================================
\section{Identifying Higher-Order Interactions from Data}
\label{sec:inference}
%%% ============================================================

The previous sections ask what dynamics a higher-order interaction produces.
Here we ask the inverse question. Given data, what can be established about the
interactions that generated them?

We divide this question into tasks that differ in how much they claim and in how
much they assume. The first is to detect higher-order \emph{dependence}, using
measures that require only the observed states
(Section~\ref{sec:inf:measures}). The second is to fit an explicit model and ask
whether higher-order terms are justified. This requires working with a model class and
raises the question of whether competing models can be told apart at all
(Section~\ref{sec:inf:identify}). The third is to recover the interaction
structure itself. This task returns the most and assumes the most
(Section~\ref{sec:inf:methods}). We treat the first at greater length because distinguishing higher-order dependence from higher-order interaction is central to this review and requires careful definition of the measures involved.

\subsection{Information-Theoretic Measures of Higher-Order Dependence}
\label{sec:inf:measures}

The key question of higher-order dependence is whether a group of three or more variables can share statistical dependence that is invisible between pairs of variables. Information theory is a natural framework to address this question by interpreting observations as outcomes of random variables \citep{Cover-Thomas_InformationTheoryBook}. We first review the tools used to investigate dependence between pairs of random variables and then discuss multivariate measures for three or more variables.

All the measures used here are built from the Shannon entropy~\citep{Cover-Thomas_InformationTheoryBook}. For a discrete
random variable $X$ with distribution $p(x)$, whose realizations are the states
$s_i$ of Section~\ref{sec:reps}, the entropy
\begin{equation}
  \mathcal{H}(X) \;=\; -\sum_{x} p(x)\log_2 p(x)
  \label{eq:entropy}
\end{equation}
measures, in bits, the uncertainty about the value of $X$, or equivalently our
ignorance of it before it is observed. Removing ignorance and gaining information are the same thing here, and we use the former reading throughout this section. We treat variables as discrete
throughout. For continuous variables the entropy is
replaced by the differential entropy, which depends on the
coordinates chosen, so the reading of Eq.~\eqref{eq:entropy} as ignorance does not
carry over directly. The mutual information, total correlation and O-information are
differences of entropies in which this dependence cancels, making them invariant
under invertible transformations of individual variables, although estimating them
from data requires density estimation or parametric, typically Gaussian, assumptions and is considerably
harder~\citep{Cover-Thomas_InformationTheoryBook}.

Suppose we observed two quantities $X$ and $Y$. The conditional entropy
$\mathcal{H}(X\mid Y) = \mathcal{H}(X,Y)-\mathcal{H}(Y)$ is the ignorance about
$X$ that survives once $Y$ is known. The reduction in ignorance of one of the quantities when the other is observed is given by the \emph{mutual information}
\begin{equation}
  I(X;Y) \;=\; \mathcal{H}(X) + \mathcal{H}(Y) - \mathcal{H}(X,Y)
  \;=\; \mathcal{H}(X) - \mathcal{H}(X\mid Y) \;=\; \mathcal{H}(Y) - \mathcal{H}(Y\mid X).
  \label{eq:mutualinfo}
\end{equation}
$I(X;Y)$ is symmetric, non-negative, and vanishes precisely when $X$ and $Y$ are independent. For time
series, the corresponding directed quantity is the \emph{transfer
entropy}~\citep{Schreiber2000}, the reduction in ignorance about the next
state of $X$ provided by the recent past of $Y$, \emph{beyond} what the recent
past of $X$ already supplies.

Moving onto three or more variables, the following example demonstrates that
higher-order dependence is possible even in the absence of pairwise dependence.
Let $X_1$ and $X_2$ be independent bits, each taking the values $0$ and $1$ with
probability $1/2$, and let $X_3 = X_1 \oplus X_2$, where $\oplus$ denotes
addition modulo two, so that $X_3=1$ when exactly one of $X_1$ and $X_2$ equals
$1$ and $X_3=0$ otherwise. Observing
$X_1$ alone leaves $X_3$ equally likely to be $0$ or $1$, because $X_2$ remains
unknown, and the same holds for every other pair, so all pairwise mutual
informations vanish, $I(X_i;X_j)=0$. The three variables are nonetheless
completely constrained, since any two of them determine the third.

The \emph{partial information decomposition} (PID) of \citet{WilliamsBeer2010}
provides a systematic framework to understand statistical dependence in multivariate systems. In its simplest case, with two sources $X_1,X_2$ and
a target $Y$, the information the sources jointly carry about the target splits
into four parts: information available from either source alone ($\mathrm{Red} $: redundant),
from one source only ($\mathrm{Unq}_1$: unique to $X_1$, $\mathrm{Unq}_2$: unique to $X_2$), and from the two
together but from neither separately ($\mathrm{Syn}$: synergistic). These parts are constrained by
\begin{equation}
  I(X_1;Y) = \mathrm{Red} + \mathrm{Unq}_1, \qquad
  I(X_2;Y) = \mathrm{Red} + \mathrm{Unq}_2, \qquad
  I(X_1,X_2;Y) = \mathrm{Red} + \mathrm{Unq}_1 + \mathrm{Unq}_2 + \mathrm{Syn}.
  \label{eq:pid}
\end{equation}
For the parity example, the dependence is purely synergistic: $\mathrm{Red}=\mathrm{Unq}_1=\mathrm{Unq}_2=0$ and
$\mathrm{Syn}=1$ bit. Since Eq.~\eqref{eq:pid} gives three equations for four unknowns, the decomposition requires an additional constraint supplied by hand, and the many inequivalent choices proposed disagree on concrete systems (\citealp{Bertschinger2014}, and others). Some properties asked of a redundancy measure are mutually inconsistent, so no agreed decomposition exists for general $N$~\citep{Gutknecht2021}. The number of parts also grows faster than exponentially with the number of sources, so PID remains practical only for small groups.

The \emph{O-information}~\citep{Rosas2019} is an alternative approach which trades the fine structure of PID for scalability. Unlike PID, it treats the variables symmetrically, requiring no division into sources and a target. Given a set of $N$ random variables, $\mathbf{X}=(X_1,\ldots,X_N)$, and denoting the collection with
$X_i$ removed as $\mathbf{X}_{-i}$, it is defined as follows:
\begin{equation}
  \Omega(\mathbf{X}) \;=\; \mathrm{TC}(\mathbf{X}) - \mathrm{DTC}(\mathbf{X}),
  \label{eq:oinfo}
\end{equation}
where
\begin{equation}
  \mathrm{TC}(\mathbf{X}) = \sum_{i} \mathcal{H}(X_i) - \mathcal{H}(\mathbf{X}),
  \qquad
  \mathrm{DTC}(\mathbf{X}) = \mathcal{H}(\mathbf{X})
     - \sum_{i}\mathcal{H}(X_i \mid \mathbf{X}_{-i}),
  \label{eq:tcdtc}
\end{equation}
are the \emph{total correlation}~\citep{Watanabe1960} and the \emph{dual total correlation}~\citep{Han1978}. Both admit a reading in terms of ignorance. $\mathrm{TC}$ is the amount by which describing each variable separately, with ignorance $\sum_i \mathcal{H}(X_i)$, overstates the joint ignorance $\mathcal{H}(\mathbf{X})$, and so measures the total constraint among the variables. $\mathrm{DTC}$ is the joint ignorance in excess of $\sum_i \mathcal{H}(X_i \mid \mathbf{X}_{-i})$, the ignorance private to each variable once all the others are known, and so measures the uncertainty that can be resolved only by knowing the configuration collectively.

$\Omega$ thus compares the ignorance resolvable by considering the parts separately with that resolvable only by considering them jointly, so that $\Omega>0$ signals a
redundancy-dominated system and $\Omega<0$ a synergy-dominated one. Three
identical copies of a bit give $\mathrm{TC}=2$, $\mathrm{DTC}=1$ and
$\Omega=+1$ bit, while the parity example gives $\mathrm{TC}=1$,
$\mathrm{DTC}=2$ and $\Omega=-1$ bit. Because
$\Omega$ is a single number for the whole group, it does not say which variables
are responsible for the balance it reports. The \emph{gradients of
O-information} recover this by measuring how $\Omega$ changes when one variable
is removed from the group, so that each variable can be labelled as contributing
redundancy or synergy~\citep{Scagliarini2023}.

O-information is the central measure used by \citet{Env2026} to show that both redundant and synergistic dependencies are possible for a system of variables coupled to a common stochastic environment with no direct interactions. The same work proves a no-go theorem stating that a time-\emph{independent} coupling to a Gaussian shared environment only produces redundancy. Synergistic dependence requires either a time-dependent environmental coupling or an interplay between the shared environment and genuine direct interactions.

All of the multivariate measures above are defined from equal-time statistics.
The \emph{dynamical O-information}~\citep{Stramaglia2021} and the \emph{total dynamical O-information}~\citep{Robiglio2025} extend them to \emph{dynamical} dependence in time series by conditioning on the past, as transfer entropy extends the mutual information, with the same sign convention.

The detection problem can be framed as a mapping between \emph{mechanisms}, the interaction and
dynamical rules governing a system, and \emph{behaviours}, the observable
statistical interdependencies that these measures quantify
\citep{Rosas2022,Robiglio2025,Marinazzo2025}. The mapping is
many-to-many: distinct mechanisms can produce identical behaviours, and simple
mechanisms can produce complex behaviours. In particular, a higher-order feature of these measures, such as synergy, does
not imply a higher-order mechanism. A triangle of
three spins with purely \emph{pairwise} antiferromagnetic couplings provides a neat example in which geometric
frustration produces an interdependence among the three spins that the sum of its pairwise interdependencies does not account for \citep{Matsuda2000,Rosas2022}, although the maximum-entropy model fixed by its pairwise marginals reproduces it exactly, so that its statistical order is two, as for any equilibrium distribution with only pairwise couplings~\citep{Schneidman2003}.

The many-to-many mapping complicates detection, but \citet{Robiglio2025} find that genuine group interactions generate stronger synergistic dynamical O-information than pairwise cliques or random triplets. The signal is invisible to low-order observables, yet appears only in a finite, system-dependent parameter region and can be masked by collective ordering.

All of these measures, dynamical ones included, are functionals of the joint distribution and its marginals. They describe how information is distributed across a group and are silent on the mechanism that produced that distribution.

\subsection{Inference and Identifiability}
\label{sec:inf:identify}

Suppose one wishes not merely to measure dependence but to fit an explicit
model and ask whether higher-order terms exist.
A standard tool is the maximum-entropy
method~\citep{Jaynes1957,Cover-Thomas_InformationTheoryBook}, which selects the distribution of greatest entropy
among those consistent with a chosen set of measured statistics, and so assumes
no more than the data require.
%The maximum-entropy distribution consistent with statistics up to order $k$ has Boltzmann form with couplings up to order $k$, so matching firing rates and pairwise correlations yields an Ising model and the test becomes one about interaction order rather than about statistics alone.

Using this method, \citet{Schneidman2006} found that for populations of about ten neurons in the
vertebrate retina a maximum-entropy model matching firing rates and pairwise correlations, an Ising model, suffices. This became a touchstone for the sufficiency of
pairwise descriptions.
However, subsequent work found that, as the population size
grows, pairwise models systematically fail, and higher-order constraints, such as the
distribution of the total activity across the population, become necessary to fit the data
\citep{Tkacik2014,Ganmor2011}. Whether a higher-order model is \emph{justified}
is therefore not absolute but depends on system, scale, and the observables one
demands the model reproduce. This is the relativity to an admissible class of
models and to a target of comparison that subquestions (i) and (ii) of
Section~\ref{sec:reps} make explicit.

Two obstacles limit how far such inference can be pushed. The first is
statistical: estimating higher-order dependence requires resolving the joint
statistics of large groups, and the data required grow exponentially with the
order, so higher-order coupling estimates are high-variance and prone to
sampling artefacts~\citep{Roudi2009}. The second is structural: reconstruction
of the interaction structure from low-order marginals is ill-posed in general,
as many distinct joint distributions are consistent with the same measured
marginals. Model selection (penalizing complexity through cross-validated
likelihood, Bayesian evidence, or description length) mitigates but does not
eliminate this, because the competing models can be genuinely
\emph{indistinguishable} given the available data.
This is the statistical
side of the reducibility and representation
question~\citep{Peixoto2026,Rommens2026,Llabres2026}. A hyperedge and a suitably
parametrized set of pairwise interactions can fit the same observations, so data
alone cannot decide between them without interventions or further structural
assumptions.
Section~\ref{sec:det:eco} gave an ecological instance, in which effective
pairwise models are fitted to abundance time series as accurately as
higher-order ones~\citep{CallejaSolanas2026}.
Parsimony then favours the
lower-order representation by default, so a positive case for higher-order
interactions must show that the added representational cost buys something
beyond goodness of fit.

\subsection{Reconstructing Higher-Order Structure}
\label{sec:inf:methods}

The third task is the most ambitious as it attempts to recover the interaction structure
itself. It can be addressed at several levels, each returning a stronger result
at the cost of stronger assumptions.

The least committal are \emph{descriptive summaries}. \citet{Santoro2023} map the evolving higher-order dependencies of a multivariate time series into topological descriptors that distinguish dynamical regimes, for example in coupled chaotic maps, which pairwise analyses miss. Such summaries serve detection and classification without identifying coupling terms.

One step further are \emph{local dynamical couplings}. Nonlinear time-series
methods, widely used in ecology, reconstruct the dynamics from observed abundances
without assuming a model and estimate, at each observed state, how strongly a change
in one variable affects the growth of another, that is, the local Jacobian element
$\partial\dot{x}_i/\partial x_j$~\citep{Kawatsu2021,Kawatsu2024}. Such a coupling is a
property of the dynamics at a given state rather than a mechanism, and it can be
present without a direct trophic link or absent despite one~\citep{Kawatsu2021}. It
is also pairwise in form by construction. For the higher-order Lotka--Volterra
model~\eqref{eq:glv},
\[
\frac{\partial\dot{x}_i}{\partial x_j}=x_i\Bigl(\alpha_{ij}+\sum_k(\beta_{ijk}+\beta_{ikj})\,x_k\Bigr),
\]
so a three-body coefficient appears only as a state dependence of a pairwise effect,
as $K_{\mathrm{eff}}(R)$ does in Section~\ref{sec:det:sync}. These methods therefore
establish a dynamical coupling without determining the mechanism that generates it.

An \emph{explicit interaction structure} can be reconstructed from dynamics. \citet{Malizia2024} recover pairwise and higher-order couplings as hypergraphs or simplicial complexes, while \citet{Wang2022} reconstruct $2$-simplicial complexes from binary dynamics. \citet{Tang2026} use sparse Bayesian regression to obtain posterior uncertainty over hyperedges and test whether the data support higher-order signal beyond a pairwise model. All such guarantees are conditional on the assumed dynamical class and data being informative enough for identification.

A \emph{statistically supported structure} is the strongest output, obtained by treating higher-order interactions as latent and asking whether the data support them. \citet{YoungPetriPeixoto2021} infer latent hyperedges from pairwise observations, including higher-order structure only when the evidence exceeds what a parsimonious pairwise model would require, which ties identification to the model-selection and representation questions of Section~\ref{sec:synthesis}. For temporal networks, \citet{Yao2021} test the evolution of triplets against a pairwise-only null and use the same triplet dynamics to improve link prediction.

All three tasks are affected by two data-related factors. First, data usually record the states of the elements rather than their interactions. Even when group membership is recorded, as in co-authorship or co-presence data, the groups are co-occurrences, and deciding which are meaningful requires a null model. \citet{Musciotto2021} filter a hypergraph by retaining the hyperlinks over-expressed relative to a random null, separating informative higher-order connections from those expected by chance.

Second, observations are often affected by measurement error, and treating a noisy observation as the structure itself propagates the error into every conclusion about interaction order. For pairwise networks, a posterior distribution over structures is inferred in place of a single estimate~\citep{Newman2018}. The higher-order counterpart is more demanding, since the same uncertain pairwise observations may be explained by a hyperedge or by several independent edges. \citet{Lizotte2023} infer a distribution over hypergraphs consistent with imperfect and indirect pairwise measurements, reconstructing empirical and synthetic systems more accurately than an equivalent model without higher-order interactions. Noise is therefore one of the reasons the estimate of interaction order is ambiguous in the first place.

The three tasks provide progressively stronger outputs, from dependence to a statistically supported structure, but none by itself distinguishes a genuine higher-order mechanism from the indirect routes of Section~\ref{sec:indirect}. Identification from data therefore establishes whether a higher-order description is supported within a model class, while the mechanism generally requires additional evidence.

\section{Discussion}

\label{sec:synthesis}

The examples reviewed above show that three questions should be distinguished when analysing higher-order interactions: \emph{what is the mechanism, what behaviour does it produce, and is a higher-order description necessary to represent either?} A behaviour marked by a higher-order dependence may result from an explicit group interaction, emerge indirectly through hidden variables or a shared environment, or arise from the description itself, through a change of variables or the aggregation of observations in time. An important conclusion of the review is therefore that higher-order mechanism, behaviour and representation are related but distinct notions.

\subsection{Mechanisms, behaviours and reducibility}

\label{sec:disc:formal}

Across the systems reviewed here, the relation between mechanism and behaviour is many-to-many, in the sense set out by \citet{Rosas2022}, \citet{Marinazzo2025} and \citet{Robiglio2025} (Section~\ref{sec:inf:measures}). Distinct mechanisms can generate the same behaviour, so an observed behaviour seldom identifies the rule that produced it. Conversely, a higher-order mechanism has no universal behavioural consequence: its effect depends on the form and sign of the interaction, the variables retained, the structure of the system and the scale at which behaviour is observed.

Higher-order dependence can also arise without an explicit many-body term. Eliminating variables generates effective higher-order interactions among those that remain, whether the eliminated variables are unobserved, as for a shared environment or latent driver (Section~\ref{sec:indirect}), or physical but omitted, as for the resources of a consumer--resource system (Section~\ref{sec:det:eco}). Temporal aggregation produces the appearance of group interactions from interactions that are pairwise at each instant. The interaction order attributed to a system can therefore change with the variables, timescales or observational resolution.

These observations motivate a more precise use of \emph{reducibility}, which asks whether a mechanism or behaviour represented by a higher-order model can be reproduced by a lower-order description. The answer depends on the admissible models, on the target, which ranges from the microscopic rule or full dynamics to a macroscopic observable or a qualitative feature such as the order of a transition, and on the tolerance, which may be exact, asymptotic or approximate.

These choices can be formalized compactly. Let $M^\star$ denote the model whose reducibility is being considered, let $\mathcal{M}$ be the class of admissible reduced models, and let $\Phi$ map a model to the quantity or property that the reduced model is required to reproduce. If $d$ is a distance or discrepancy measure and $\epsilon\geq0$ is the accepted tolerance, we say that $M^\star$ is reducible to $\mathcal{M}$ with respect to $(\Phi,d,\epsilon)$ if

\begin{equation}
\exists\, M\in\mathcal{M}\ \text{such that}\quad
d\bigl(\Phi(M),\Phi(M^\star)\bigr)\leq\epsilon.
\label{eq:reducibility}
\end{equation}

The target $\Phi$ is deliberately general. For equilibrium systems it may be a distribution or one of its observables; for dynamics it may be a generator, transition kernel or trajectory distribution; and for phenomenological comparisons it may be a scalar signature such as a critical point or transition type. Enlarging $\mathcal{M}$, weakening $\Phi$, or increasing $\epsilon$ makes reduction easier. Thus, a model that is irreducible to pairwise interactions on its original variables may become reducible when auxiliary variables are admitted, a microscopic rule may be irreducible even when a macroscopic observable is reproducible, and a full dynamical description may be irreducible even though a qualitative transition is shared with a lower-order model. Algebraic equivalence is therefore not sufficient for a reduction to be physically meaningful. Admitting auxiliary variables or state-dependent rules can remove explicit higher-order terms while moving their complexity into the state space or the rule (Sections~\ref{sec:det:sync} and~\ref{sec:stoch:statmech}). Whether a reduction is meaningful then depends on context. It may lower the cost of the complete description, structure and rule together~\citep{Rommens2026,Rommens2026b}, or its additional variables may correspond to measurable quantities, as the resources of Section~\ref{sec:det:eco} do and the auxiliary variable of Section~\ref{sec:stoch:statmech} does not.

The distinction between mechanism and behaviour is therefore expressed naturally through the target $\Phi$. A \emph{mechanistic} target retains information about the governing law, such as a Hamiltonian, generator, vector field or transition rule. In contrast, a \emph{behavioural} target concerns only selected consequences of that law, so reproducing it is a weaker requirement. In equilibrium these may be functionals of the joint distribution $p_N(\mathbf{s})$, including macroscopic observables and information-theoretic measures. Dynamical behaviour generally requires a richer object, such as a transition kernel or distribution over trajectories, because a stationary distribution alone does not determine the dynamics that produced it. Two models with the same stationary distribution may differ in probability currents, entropy production or temporal correlations, which are then targets in their own right. In open and functionally organised systems the relevant invariants may instead be constraints on admissible trajectories, a perspective developed independently of interaction order~\citep{Monroy2026a,Monroy2026b}.

Criterion~\eqref{eq:reducibility} concerns models alone, whereas identifiability concerns what a given body of data can establish about the model that generated it (Section~\ref{sec:inf:identify}). The two are distinct: a reduction may be exact for a macroscopic observable that the data do not measure, and data may fail to separate models that are not reductions of one another. Irreducibility and non-identifiability are therefore different conclusions.

\subsection{Representation and mechanistic equivalence}

\label{sec:synthesis:degeneracy}

The structural and variable representations introduced in Section~\ref{sec:reps} both bear on whether a higher-order description is necessary. A mechanism that is higher-order among one set of variables may be pairwise among another.

The $p$-spin construction is the concrete case for the variable representation (Section~\ref{sec:stoch:statmech}). Replacing a $p$-body factor by pairwise couplings to an auxiliary variable represents the same dependence in an extended variable space rather than the same mechanism on the same variables, so whether the interaction counts as reducible depends on whether that extension is admitted~\citep{Rommens2026}. The structural representation acts similarly, since projecting a hypergraph onto a graph discards information about which groups interacted jointly~\citep{Peixoto2026,LaRockLambiotte2025}.

Three forms of equivalence are useful here. \emph{Representational equivalence} asks whether two mathematical descriptions encode the same relevant model or dynamics. \emph{Inferential equivalence} asks whether the available observations can distinguish the descriptions. \emph{Mechanistic equivalence} asks whether they correspond to the same underlying mechanism. The first can often be established mathematically, the second statistically, whereas the third generally requires additional physical, biological or causal information. These distinctions matter most for indirect mechanisms, and they explain why a claim of reducibility must specify both the variable representation admitted and the aspect of the system to be preserved.

\subsection{Common patterns across mechanisms and behaviours}

\label{sec:disc:patterns}

Several patterns recur across the examples.

First, whether a higher-order contribution is genuinely necessary often depends on whether its effect is additive or separable into lower-order terms. Linear consensus and voter dynamics can reduce exactly to pairwise descriptions (Section~\ref{sec:det:sync}), whereas genuinely non-additive group dependence generally has to be retained somewhere in the representation~\citep{Neuhauser2020,Sahasrabuddhe2021,Llabres2026,Xie2026}. Non-additivity is therefore more directly relevant to mechanistic irreducibility than the presence of a hyperedge or of a fitted higher-order coefficient.

Second, interaction order often acts as a control parameter rather than a binary switch. In the $p$-spin model (Section~\ref{sec:stoch:statmech}), increasing $p$ progressively changes the covariance structure of energies and the organization of the solution space~\citep{Derrida1980}. In ecological models, interaction order can alter the scaling and even the sign of diversity--stability relationships~\citep{Bairey2016}. Random $k$-satisfiability shows the opposite case, a genuine threshold at the pairwise boundary, since satisfiability is decidable in polynomial time for $k=2$ and NP-complete for every $k\geq3$~\citep{MezardMontanari2009}. Whether interaction order reshapes a system progressively or at a threshold is therefore itself a finding rather than something to assume.

Third, a higher-order mechanism has no universal behavioural consequence. Ecology illustrates this particularly clearly (Section~\ref{sec:det:eco}). Higher-order terms favour coexistence in some models~\citep{Grilli2017,Bairey2016} and impair it in others~\citep{Gibbs2022}, with the outcome depending on abundance normalization, the distribution of higher-order coefficients, their signs and their correlation with pairwise interactions~\citep{Gibbs2024}. Introducing a higher-order term therefore does not by itself determine the direction of a behavioural change.

Fourth, the phenomenological signatures most often associated with higher-order models are frequently \emph{non-diagnostic} in the sense of Section~\ref{sec:reps}, reproduced by a lower-order model that does not capture the original mechanism. Unlike higher-order statistical dependence, which has a well-defined order, these signatures are not themselves higher-order. Discontinuous transitions, bistability and hysteresis occur in higher-order synchronization and contagion models (Sections~\ref{sec:det:sync} and~\ref{sec:stoch:contagion}), but can also arise in systems represented by graphs, through heterogeneity, nonlinear response functions or feedback. Period doubling and chaos are likewise generic to nonlinear dynamics. Such signatures may motivate a search for higher-order structure, but they do not by themselves demonstrate its necessity.

Finally, the relevant interaction order can depend on scale. Coarse-graining, elimination of variables, finite observation windows and temporal aggregation can create, remove or transform effective interactions~\citep{Lucas2026,Meloni2026}. Higher-order structure should therefore be interpreted relative to the variables, timescale and resolution used to define the system.

\subsection{When is higher-order structure necessary?}

\label{sec:synthesis:cases}

The cases reviewed above show that claims of necessity are meaningful only relative to a specified target and comparison class. Table~\ref{tab:verdicts} summarizes representative comparisons between mechanism, behaviour and lower-order alternatives.

\begin{table}[htbp]
\caption{Mechanisms, behaviours and reducibility across the models reviewed. The table distinguishes the mechanism, which is not higher-order in every case, from the behaviour or target used for comparison. ``Irreducible'' denotes failure of reduction within the stated lower-order class; ``reducible'' denotes an exact or explicitly demonstrated reduction; ``non-diagnostic'' denotes a behaviour or observable reproduced by a lower-order model without reproducing the original mechanism; ``non-identifiable'' denotes a case in which the available data do not distinguish the higher-order model from a lower-order alternative, a conclusion about the evidence rather than about the existence of a reduction; and ``open'' denotes cases for which an appropriate comparison remains to be carried out. Where a verdict is split, the qualifier indicates the aspect to which it applies, so that ``irreducible (mechanism); reducible (structure)'' means that the mechanism has no additive counterpart although a graph suffices to express it.}
\label{tab:verdicts}
\footnotesize
\setlength{\tabcolsep}{4pt}
\begin{tabular}{p{2.3cm} p{3.1cm} p{3.1cm} p{3.1cm} p{2.3cm}}
\toprule
\textbf{Case} & \textbf{Mechanism} & \textbf{Behaviour / target} &
\textbf{Lower-order comparison} & \textbf{Conclusion for this target} \\
\midrule
Explosive synchronization
& Higher-order group coupling
& Discontinuous transition and hysteresis
& Pairwise models with quenched degree--frequency correlations or frequency-weighted coupling
& Non-diagnostic \\ \midrule
Higher-order Kuramoto, all-to-all
& Triadic and tetradic phase coupling
& Macroscopic evolution of $R$
& Pairwise Kuramoto model with state-independent coupling
& Irreducible \\ \midrule
Structured higher-order synchronization
& Link, triangle and higher-order simplex couplings
& Linear stability of synchrony; full nonlinear dynamics
& Rescaled pairwise coupling (stability); constructed pairwise models (dynamics)
& Reducible (stability); open (dynamics) \\ \midrule
Simplicial / hypergraph contagion
& Group-dependent transmission
& Bistability and discontinuous threshold
& Graph models with additive, independent pairwise transmission
& Irreducible (mechanism); reducible (structure) \\ \midrule
Synergistic contagion
& Non-additive group transmission
& Microscopic transmission rule
& Pairwise model with fixed transmission rates
& Irreducible \\ \midrule
Triadic percolation
& Group-dependent activation
& Dynamical connectivity
& Pairwise percolation with fixed edges
& Open \\ \midrule
Consumer--resource competition
& Pairwise consumer--resource interactions
& Effective higher-order consumer dynamics
& Pairwise model retaining the resources explicitly
& Reducible (enlarged variables) \\ \midrule
Higher-order Lotka--Volterra fitted to data
& Direct higher-order competition coefficients
& Abundance time series to a stated tolerance
& Pairwise model with nonlinear functional responses
& Non-identifiable \\ \midrule
Cooperation on higher-order structures
& Multiplayer group-dependent payoffs
& Explosive transition to cooperation and strategy coexistence
& Pairwise games on structured populations
& Non-diagnostic \\ \midrule
$p$-spin glass, $p\geq3$
& $p$-body Hamiltonian
& $p$-body interaction structure on the original spins
& Pairwise Hamiltonians on the original $N$ spins
& Irreducible \\ \midrule
$p$-spin factor
& $p$-body equilibrium factor
& Equilibrium distribution of observed spins
& Pairwise representation with auxiliary variables
& Reducible \\
\bottomrule
\end{tabular}
\end{table}

In synchronization (Section~\ref{sec:det:sync}), the reduced amplitude equation contains a configuration-dependent effective coupling, which is not a pairwise mechanism, yet explosive synchronization is also reproduced by pairwise models with suitable heterogeneity. The two statements concern different targets, the microscopic mechanism and a qualitative collective behaviour.

In ecology (Section~\ref{sec:det:eco}), eliminating resources generates higher-order Lotka--Volterra coefficients among consumers whose underlying interactions are pairwise, so the higher-order description is reducible if the resources may be restored. An effective pairwise model of the observed consumers may also reproduce the abundance dynamics to a specified accuracy~\citep{CallejaSolanas2026}, a different reduction with a different model class and target.

For the $p$-spin model (Section~\ref{sec:stoch:statmech}), a pairwise Hamiltonian on the original spins cannot reproduce the $p$-body structure, whereas auxiliary variables reproduce the equilibrium factor exactly. The higher-order dependence is then carried by additional degrees of freedom, so reducibility depends on whether the enlarged representation is admitted.

Contagion (Section~\ref{sec:stoch:contagion}) separates the two representational choices of Section~\ref{sec:reps}. A susceptible--infected process on a hypergraph coincides with one on the projected graph exactly when group transmission is linear in the number of infected members~\citep{Xie2026}, so nonlinearity in that count carries the mechanistic content, and a synergistic rule on a graph supplies it as readily as a filled triangle does. The structural representation is therefore reducible, since a graph suffices, whereas the mechanism is not, since the rule remains non-additive. Bistability and discontinuous thresholds follow from the non-additivity rather than from the architecture, which is why they do not discriminate between the two.

Triadic percolation reducibility (Section~\ref{sec:stoch:contagion}) remains open for a different reason. Pairwise percolation with fixed edges cannot produce a time-dependent giant component, but period doubling and chaos are generic to nonlinear dynamics. The triadic update depends on the global giant component rather than on a local rule, so the comparisons used elsewhere in this review do not apply directly.

Taken together, these cases suggest that the strongest evidence for the necessity of a higher-order mechanism is not a striking collective behaviour alone. It requires showing that plausible lower-order alternatives do not reproduce a clearly specified target, after accounting for admissible changes of representation, hidden variables and observational scale. Conversely, reproducing a behaviour with a lower-order model establishes only that this particular target does not require the higher-order mechanism within the tested class.

\subsection{What can data establish?}

\label{sec:disc:inference}

The methods of Section~\ref{sec:inference} establish dependence, and at best a structure consistent with the data under an assumed dynamics. However, they do not by themselves identify mechanisms without additional structural assumptions, sufficiently resolved dynamics, or interventions (Section~\ref{sec:indirect}). Some mechanistic discrimination is nevertheless possible.

The no-go theorem by \citet{Env2026} shows that non-interacting variables coupled to a single Gaussian shared environment through a common time profile can display redundancy but not synergy. Observed synergy therefore rules out this mechanism as a complete explanation, pointing instead to heterogeneous couplings, direct interactions, or more complex environments.
Results of this kind, which exclude specific indirect mechanisms rather than confirm a direct one, are a realistic target for observational work.

\subsection{Open problems}

\label{sec:outlook}

A central open problem is to connect governing mechanisms to the statistical dependence and phenomenological signatures that can actually be measured. For equilibrium systems, the joint distribution $p_N(\mathbf{s})$ provides a natural probabilistic description; for dynamical systems, the corresponding object is more generally a transition law or probability distribution over trajectories. Developing a framework that connects these descriptions to interaction structure, collective behaviour and the reducibility criterion of Eq.~\eqref{eq:reducibility} would help place the diverse approaches reviewed here on a common footing.

A related question is whether higher-order interactions define genuinely distinct large-scale behaviour or become irrelevant under coarse-graining. Two of the cases reviewed here point in opposite directions. In the $p$-spin glass (Section~\ref{sec:stoch:statmech}), increasing the interaction order changes the thermodynamics qualitatively, splitting the dynamical and thermodynamic transitions in a way that no pairwise Hamiltonian on the same spins reproduces~\citep{Derrida1980}. In simplicial contagion (Section~\ref{sec:stoch:contagion}), by contrast, higher-order simplices can become irrelevant under renormalization, the coarse-grained dynamics taking the same form as synergistic or threshold contagion on a graph~\citep{Meloni2026}. \citet{Villegas2026} argue that this is a relevant test of novelty, since collective behaviour is classified by the ingredients that survive at long scales and interaction order alone is not a universality label. A systematic connection between interaction reduction and renormalization-group flow could clarify when higher-order structure survives at macroscopic scales and when it is replaced by effective lower-order descriptions.

A third open problem is to establish what would justify the step from dependence to mechanism. Identifiability theory~\citep{Villaverde2019} offers one route, by characterizing in advance the equivalence classes of models that generate the same observables, so that the limits of what the data could determine are known before fitting. A second is to seek exclusion results rather than confirmations, as in the shared-environment result of Section~\ref{sec:disc:inference}, where a restricted class of indirect mechanisms is shown to produce only a restricted kind of statistical dependence. Constructing comparable results for variable elimination and temporal aggregation would turn each indirect route into a hypothesis that observations can reject. Where intervention is possible, perturbing a candidate group remains the most decisive evidence, while expressing reducibility in causal terms would clarify what such an intervention can establish.

Beyond-mean-field analysis is also needed. Many reductions rely on all-to-all, annealed or well-mixed approximations that suppress local correlations and fluctuations. On sparse or structured substrates, closure relations may fail and local higher-order motifs may remain dynamically relevant even when the corresponding mean-field model is reducible. Determining which reductions survive finite-size and structural fluctuations is therefore an important challenge.

\subsection{Conclusion}

Taken together, the review suggests that the question of whether higher-order interactions are ``necessary'' is too broad to have a single answer. Statistical mechanics connects interaction order to the organization of state space, dynamical-systems theory connects it to mechanisms and collective behaviour, and information theory quantifies multivariate dependence without fixing a particular representation. Bringing these perspectives together replaces the abstract question of necessity with a more precise one: \emph{which mechanisms generate which behaviours, which lower-order models reproduce a specified target, at what scale and on what evidence?} This framing does not diminish the importance of higher-order interactions. Rather, it identifies the circumstances in which their mechanisms, behaviours and necessity are mathematically distinguishable and empirically testable.

\section*{Acknowledgements}

The author used Claude (Anthropic) to improve the English language, clarity and
readability of this manuscript. All content was subsequently reviewed and edited
by the author.

\section*{Funding}

This research was funded by the BBSRC grant no. BB/Y513027/1.

\section*{Data availability}

This article is a review, and no new datasets were generated or analysed.


\begin{thebibliography}{115}
\expandafter\ifx\csname natexlab\endcsname\relax\def\natexlab#1{#1}\fi
\expandafter\ifx\csname urlstyle\endcsname\relax
  \expandafter\ifx\csname doi\endcsname\relax
  \def\doi#1{doi:\discretionary{}{}{}#1}\fi \else
  \expandafter\ifx\csname doi\endcsname\relax
  \def\doi{doi:\discretionary{}{}{}\begingroup \urlstyle{rm}\Url}\fi \fi
\expandafter\ifx\csname selectlanguage\endcsname\relax
  \def\selectlanguage#1{}\fi

\bibitem[{Battiston et~al.(2020)Battiston, Cencetti, Iacopini, Latora, Lucas,
  Patania et~al.}]{Battiston2020}
Battiston F, Cencetti G, Iacopini I, Latora V, Lucas M, Patania A, et~al.
\newblock Networks beyond pairwise interactions: Structure and dynamics.
\newblock {\em Physics Reports\/} {\bf 874} (2020) 1--92.
\newblock \doi{10.1016/j.physrep.2020.05.004}.

\bibitem[{Battiston et~al.(2021)Battiston, Amico, Barrat, Bianconi, Ferraz~de
  Arruda, Franceschiello et~al.}]{Battiston2021}
Battiston F, Amico E, Barrat A, Bianconi G, Ferraz~de Arruda G, Franceschiello
  B, et~al.
\newblock The physics of higher-order interactions in complex systems.
\newblock {\em Nature Physics\/} {\bf 17} (2021) 1093--1098.
\newblock \doi{10.1038/s41567-021-01371-4}.

\bibitem[{Bick et~al.(2023)Bick, Gross, Harrington, and Schaub}]{Bick2023}
Bick C, Gross E, Harrington HA, Schaub MT.
\newblock What are higher-order networks?
\newblock {\em SIAM Review\/} {\bf 65} (2023) 686--731.
\newblock \doi{10.1137/21M1414024}.

\bibitem[{Majhi et~al.(2022)Majhi, Perc, and Ghosh}]{MajhiPercGhosh2022}
Majhi S, Perc M, Ghosh D.
\newblock Dynamics on higher-order networks: a review.
\newblock {\em Journal of the Royal Society Interface\/} {\bf 19} (2022)
  20220043.
\newblock \doi{10.1098/rsif.2022.0043}.

\bibitem[{Boccaletti et~al.(2023)Boccaletti, De~Lellis, del Genio,
  Alfaro-Bittner, Criado, Jalan et~al.}]{Boccaletti2023}
Boccaletti S, De~Lellis P, del Genio CI, Alfaro-Bittner K, Criado R, Jalan S,
  et~al.
\newblock The structure and dynamics of networks with higher order
  interactions.
\newblock {\em Physics Reports\/} {\bf 1018} (2023) 1--64.
\newblock \doi{10.1016/j.physrep.2023.04.002}.

\bibitem[{Skardal and Arenas(2020)}]{SkardalArenas2020}
Skardal PS, Arenas A.
\newblock Higher order interactions in complex networks of phase oscillators
  promote abrupt synchronization switching.
\newblock {\em Communications Physics\/} {\bf 3} (2020) 218.
\newblock \doi{10.1038/s42005-020-00485-0}.

\bibitem[{Iacopini et~al.(2019)Iacopini, Petri, Barrat, and
  Latora}]{Iacopini2019}
Iacopini I, Petri G, Barrat A, Latora V.
\newblock Simplicial models of social contagion.
\newblock {\em Nature Communications\/} {\bf 10} (2019) 2485.
\newblock \doi{10.1038/s41467-019-10431-6}.

\bibitem[{Grilli et~al.(2017)Grilli, Barab{\'a}s, Michalska-Smith, and
  Allesina}]{Grilli2017}
Grilli J, Barab{\'a}s G, Michalska-Smith MJ, Allesina S.
\newblock Higher-order interactions stabilize dynamics in competitive network
  models.
\newblock {\em Nature\/} {\bf 548} (2017) 210--213.
\newblock \doi{10.1038/nature23273}.

\bibitem[{Derrida(1980)}]{Derrida1980}
Derrida B.
\newblock Random-energy model: Limit of a family of disordered models.
\newblock {\em Physical Review Letters\/} {\bf 45} (1980) 79--82.
\newblock \doi{10.1103/PhysRevLett.45.79}.

\bibitem[{Gross and M{\'e}zard(1984)}]{GrossMezard1984}
Gross DJ, M{\'e}zard M.
\newblock The simplest spin glass.
\newblock {\em Nuclear Physics B\/} {\bf 240} (1984) 431--452.
\newblock \doi{10.1016/0550-3213(84)90237-2}.

\bibitem[{Billick and Case(1994)}]{BillickCase1994}
Billick I, Case TJ.
\newblock Higher order interactions in ecological communities: What are they
  and how can they be detected?
\newblock {\em Ecology\/} {\bf 75} (1994) 1529--1543.
\newblock \doi{10.2307/1939614}.

\bibitem[{Gokhale and Traulsen(2010)}]{Gokhale2010}
Gokhale CS, Traulsen A.
\newblock Evolutionary games in the multiverse.
\newblock {\em Proceedings of the National Academy of Sciences\/} {\bf 107}
  (2010) 5500--5504.
\newblock \doi{10.1073/pnas.0912214107}.

\bibitem[{Schneidman et~al.(2006)Schneidman, Berry, Segev, and
  Bialek}]{Schneidman2006}
Schneidman E, Berry MJ, Segev R, Bialek W.
\newblock Weak pairwise correlations imply strongly correlated network states
  in a neural population.
\newblock {\em Nature\/} {\bf 440} (2006) 1007--1012.
\newblock \doi{10.1038/nature04701}.

\bibitem[{Peixoto et~al.(2026)Peixoto, Peel, Gross, and
  De~Domenico}]{Peixoto2026}
Peixoto TP, Peel L, Gross T, De~Domenico M.
\newblock Graphs are maximally expressive for higher-order interactions.
\newblock {\em arXiv preprint arXiv:2602.16937\/}  (2026).

\bibitem[{Rommens et~al.(2026{\natexlab{a}})Rommens, Traversa, Ferraz~de
  Arruda, and Moreno}]{Rommens2026}
Rommens C, Traversa P, Ferraz~de Arruda G, Moreno Y.
\newblock The informational cost of structure: Representational complexity in
  networked dynamical systems.
\newblock {\em arXiv preprint arXiv:2607.03608\/}  (2026{\natexlab{a}}).

\bibitem[{Rosas et~al.(2022)Rosas, Mediano, Luppi, Varley, Lizier, Stramaglia
  et~al.}]{Rosas2022}
Rosas FE, Mediano PAM, Luppi AI, Varley TF, Lizier JT, Stramaglia S, et~al.
\newblock Disentangling high-order mechanisms and high-order behaviours in
  complex systems.
\newblock {\em Nature Physics\/} {\bf 18} (2022) 476--477.
\newblock \doi{10.1038/s41567-022-01548-5}.

\bibitem[{Marinazzo(2025)}]{Marinazzo2025}
Marinazzo D.
\newblock With behaviors like these in complex systems, who needs mechanisms?
\newblock {\em Physics\/} {\bf 18} (2025) 71.
\newblock \doi{10.1103/Physics.18.71}.

\bibitem[{Robiglio et~al.(2025)Robiglio, Neri, Coppes, Agostinelli, Battiston,
  Lucas et~al.}]{Robiglio2025}
Robiglio T, Neri M, Coppes D, Agostinelli C, Battiston F, Lucas M, et~al.
\newblock Synergistic signatures of group mechanisms in higher-order systems.
\newblock {\em Physical Review Letters\/} {\bf 134} (2025) 137401.
\newblock \doi{10.1103/PhysRevLett.134.137401}.

\bibitem[{Abril-Berm{\'u}dez et~al.(2026)Abril-Berm{\'u}dez, Fisher, Gramain,
  and P{\'e}rez-Reche}]{Env2026}
Abril-Berm{\'u}dez FS, Fisher DN, Gramain JB, P{\'e}rez-Reche FJ.
\newblock Environment-driven emergence of higher-order collective behavior.
\newblock {\em arXiv preprint arXiv:2602.15256\/}  (2026).

\bibitem[{Cencetti et~al.(2021)Cencetti, Battiston, Lepri, and
  Karsai}]{Cencetti2021}
Cencetti G, Battiston F, Lepri B, Karsai M.
\newblock Temporal properties of higher-order interactions in social networks.
\newblock {\em Scientific Reports\/} {\bf 11} (2021) 7028.
\newblock \doi{10.1038/s41598-021-86469-8}.

\bibitem[{Battiston et~al.(2026)Battiston, Bick, Lucas, Millán, Skardal, and
  Zhang}]{Battiston2026}
Battiston F, Bick C, Lucas M, Millán AP, Skardal PS, Zhang Y.
\newblock Collective dynamics on higher-order networks.
\newblock {\em Nature Reviews Physics\/} {\bf 8} (2026) 146--159.
\newblock \doi{10.1038/s42254-025-00916-3}.

\bibitem[{Rosas et~al.(2019)Rosas, Mediano, Gastpar, and Jensen}]{Rosas2019}
Rosas FE, Mediano PAM, Gastpar M, Jensen HJ.
\newblock Quantifying high-order interdependencies via multivariate extensions
  of the mutual information.
\newblock {\em Physical Review E\/} {\bf 100} (2019) 032305.
\newblock \doi{10.1103/PhysRevE.100.032305}.

\bibitem[{Neuh{\"a}user et~al.(2020)Neuh{\"a}user, Mellor, and
  Lambiotte}]{Neuhauser2020}
Neuh{\"a}user L, Mellor A, Lambiotte R.
\newblock Multibody interactions and nonlinear consensus dynamics on networked
  systems.
\newblock {\em Physical Review E\/} {\bf 101} (2020) 032310.
\newblock \doi{10.1103/PhysRevE.101.032310}.

\bibitem[{Llabr{\'e}s et~al.(2026)Llabr{\'e}s, V{\'a}zquez, Toral, and
  San~Miguel}]{Llabres2026}
Llabr{\'e}s J, V{\'a}zquez F, Toral R, San~Miguel M.
\newblock Reducibility of higher-order to pairwise interactions: Social impact
  models on hypergraphs.
\newblock {\em Physical Review Research\/}  (2026).

\bibitem[{Lucas et~al.(2026)Lucas, Gallo, Ghavasieh, Battiston, and
  De~Domenico}]{Lucas2026}
Lucas M, Gallo L, Ghavasieh A, Battiston F, De~Domenico M.
\newblock Reducibility of higher-order networks from dynamics.
\newblock {\em Nature Communications\/} {\bf 17} (2026) 1551.
\newblock \doi{10.1038/s41467-025-68273-4}.

\bibitem[{Xie et~al.(2026)Xie, Li, Sun, He, Zhang, and Chen}]{Xie2026}
Xie M, Li A, Sun Y, He S, Zhang ZK, Chen J.
\newblock Transformability of dynamics on higher-order networks.
\newblock {\em Communications Physics\/} {\bf 9} (2026) 149.
\newblock \doi{10.1038/s42005-026-02555-1}.

\bibitem[{Rommens et~al.(2026{\natexlab{b}})Rommens, Traversa, Ferraz~de
  Arruda, and Moreno}]{Rommens2026b}
Rommens C, Traversa P, Ferraz~de Arruda G, Moreno Y.
\newblock Beyond expressiveness in pairwise and higher-order models.
\newblock {\em arXiv preprint arXiv:2609.26031\/}  (2026{\natexlab{b}}).

\bibitem[{Amari(2001)}]{Amari2001}
Amari SI.
\newblock Information geometry on hierarchy of probability distributions.
\newblock {\em IEEE Transactions on Information Theory\/} {\bf 47} (2001)
  1701--1711.
\newblock \doi{10.1109/18.930911}.

\bibitem[{Bianconi(2021)}]{Bianconi2021book}
Bianconi G.
\newblock {\em Higher-Order Networks\/}.
\newblock Elements in the Structure and Dynamics of Complex Networks (Cambridge
  University Press) (2021).
\newblock \doi{10.1017/9781108770996}.

\bibitem[{Silk et~al.(2022)Silk, Wilber, and Fefferman}]{Silk2022}
Silk MJ, Wilber MQ, Fefferman NH.
\newblock Capturing complex interactions in disease ecology with simplicial
  sets.
\newblock {\em Ecology Letters\/} {\bf 25} (2022) 2217--2231.
\newblock \doi{10.1111/ELE.14079}.

\bibitem[{Acebrón et~al.(2005)Acebrón, Bonilla, Vicente, Ritort, and
  Spigler}]{Acebron2005}
Acebrón JA, Bonilla LL, Vicente CJ, Ritort F, Spigler R.
\newblock The kuramoto model: A simple paradigm for synchronization phenomena.
\newblock {\em Reviews of Modern Physics\/} {\bf 77} (2005) 137.
\newblock \doi{10.1103/RevModPhys.77.137}.

\bibitem[{G{\'o}mez-Garde{\~n}es et~al.(2011)G{\'o}mez-Garde{\~n}es, G{\'o}mez,
  Arenas, and Moreno}]{GomezGardenes2011}
G{\'o}mez-Garde{\~n}es J, G{\'o}mez S, Arenas A, Moreno Y.
\newblock Explosive synchronization transitions in scale-free networks.
\newblock {\em Physical Review Letters\/} {\bf 106} (2011) 128701.
\newblock \doi{10.1103/PhysRevLett.106.128701}.

\bibitem[{Zhang et~al.(2013)Zhang, Hu, Kurths, and Liu}]{Zhang2013}
Zhang X, Hu X, Kurths J, Liu Z.
\newblock Explosive synchronization in a general complex network.
\newblock {\em Physical Review E\/} {\bf 88} (2013) 010802.
\newblock \doi{10.1103/PhysRevE.88.010802}.

\bibitem[{Ott and Antonsen(2008)}]{OttAntonsen2008}
Ott E, Antonsen TM.
\newblock Low dimensional behavior of large systems of globally coupled
  oscillators.
\newblock {\em Chaos\/} {\bf 18} (2008) 037113.
\newblock \doi{10.1063/1.2930766}.

\bibitem[{Zhang et~al.(2015)Zhang, Boccaletti, Guan, and Liu}]{Zhang2015}
Zhang X, Boccaletti S, Guan S, Liu Z.
\newblock Explosive synchronization in adaptive and multilayer networks.
\newblock {\em Physical Review Letters\/} {\bf 114} (2015) 038701.
\newblock \doi{10.1103/PhysRevLett.114.038701}.

\bibitem[{Ott and Antonsen(2009)}]{OttAntonsen2009}
Ott E, Antonsen TM.
\newblock Long time evolution of phase oscillator systems.
\newblock {\em Chaos\/} {\bf 19} (2009) 023117.
\newblock \doi{10.1063/1.3136851}.

\bibitem[{Mill{\'a}n et~al.(2020)Mill{\'a}n, Torres, and
  Bianconi}]{MillanTorresBianconi2020}
Mill{\'a}n AP, Torres JJ, Bianconi G.
\newblock Explosive higher-order kuramoto dynamics on simplicial complexes.
\newblock {\em Physical Review Letters\/} {\bf 124} (2020) 218301.
\newblock \doi{10.1103/PhysRevLett.124.218301}.

\bibitem[{Lucas et~al.(2020)Lucas, Cencetti, and
  Battiston}]{LucasBattiston2020}
Lucas M, Cencetti G, Battiston F.
\newblock Multiorder laplacian for synchronization in higher-order networks.
\newblock {\em Physical Review Research\/} {\bf 2} (2020) 033410.
\newblock \doi{10.1103/PhysRevResearch.2.033410}.

\bibitem[{Gambuzza et~al.(2021)Gambuzza, Di~Patti, Gallo, Lepri, Romance,
  Criado et~al.}]{Gambuzza2021}
Gambuzza LV, Di~Patti F, Gallo L, Lepri S, Romance M, Criado R, et~al.
\newblock Stability of synchronization in simplicial complexes.
\newblock {\em Nature Communications\/} {\bf 12} (2021) 1255.
\newblock \doi{10.1038/s41467-021-21486-9}.

\bibitem[{Ren et~al.(2023)Ren, Lei, Grebogi, and Baptista}]{Ren2023}
Ren X, Lei Y, Grebogi C, Baptista MS.
\newblock The complementary contribution of each order topology into the
  synchronization of multi-order networks.
\newblock {\em Chaos\/} {\bf 33} (2023) 111101.
\newblock \doi{10.1063/5.0177687}.

\bibitem[{Skardal and Arenas(2019)}]{SkardalArenas2019}
Skardal PS, Arenas A.
\newblock Abrupt desynchronization and extensive multistability in globally
  coupled oscillator simplexes.
\newblock {\em Physical Review Letters\/} {\bf 122} (2019) 248301.
\newblock \doi{10.1103/PhysRevLett.122.248301}.

\bibitem[{Sahasrabuddhe et~al.(2021)Sahasrabuddhe, Neuh{\"a}user, and
  Lambiotte}]{Sahasrabuddhe2021}
Sahasrabuddhe R, Neuh{\"a}user L, Lambiotte R.
\newblock Modelling non-linear consensus dynamics on hypergraphs.
\newblock {\em Journal of Physics: Complexity\/} {\bf 2} (2021) 025006.
\newblock \doi{10.1088/2632-072X/abcea3}.

\bibitem[{Hofbauer and Sigmund(1998)}]{HofbauerSigmund1998}
Hofbauer J, Sigmund K.
\newblock {\em Evolutionary Games and Population Dynamics\/} (Cambridge
  University Press) (1998).
\newblock \doi{10.1017/CBO9781139173179}.

\bibitem[{Bairey et~al.(2016)Bairey, Kelsic, and Kishony}]{Bairey2016}
Bairey E, Kelsic ED, Kishony R.
\newblock High-order species interactions shape ecosystem diversity.
\newblock {\em Nature Communications\/} {\bf 7} (2016) 12285.
\newblock \doi{10.1038/ncomms12285}.

\bibitem[{Gibbs et~al.(2022)Gibbs, Levin, and Levine}]{Gibbs2022}
Gibbs T, Levin SA, Levine JM.
\newblock Coexistence in diverse communities with higher-order interactions.
\newblock {\em Proceedings of the National Academy of Sciences\/} {\bf 119}
  (2022) e2205063119.
\newblock \doi{10.1073/pnas.2205063119}.

\bibitem[{May(1972)}]{May1972}
May RM.
\newblock Will a large complex system be stable?
\newblock {\em Nature\/} {\bf 238} (1972) 413--414.
\newblock \doi{10.1038/238413a0}.

\bibitem[{May(2001)}]{May2001}
May RM.
\newblock {\em Stability and Complexity in Model Ecosystems\/} (Princeton
  University Press) (2001).

\bibitem[{Levine et~al.(2017)Levine, Bascompte, Adler, and
  Allesina}]{Levine2017}
Levine JM, Bascompte J, Adler PB, Allesina S.
\newblock Beyond pairwise mechanisms of species coexistence in complex
  communities.
\newblock {\em Nature\/} {\bf 546} (2017) 56--64.
\newblock \doi{10.1038/nature22898}.

\bibitem[{Gibbs et~al.(2024)Gibbs, Gellner, Levin, McCann, Hastings, and
  Levine}]{Gibbs2024}
Gibbs TL, Gellner G, Levin SA, McCann KS, Hastings A, Levine JM.
\newblock When can higher-order interactions produce stable coexistence?
\newblock {\em Ecology Letters\/} {\bf 27} (2024) e14458.
\newblock \doi{10.1111/ele.14458}.

\bibitem[{Mayfield and Stouffer(2017)}]{Mayfield2017}
Mayfield MM, Stouffer DB.
\newblock Higher-order interactions capture unexplained complexity in diverse
  communities.
\newblock {\em Nature Ecology \& Evolution\/} {\bf 1} (2017) 0062.
\newblock \doi{10.1038/s41559-016-0062}.

\bibitem[{Buche et~al.(2024)Buche, Bartomeus, and Godoy}]{BucheGodoy2024}
Buche L, Bartomeus I, Godoy O.
\newblock Multitrophic higher-order interactions modulate species persistence.
\newblock {\em The American Naturalist\/} {\bf 203} (2024) 458--472.
\newblock \doi{10.1086/729222}.

\bibitem[{Shen et~al.(2023)Shen, Lemmen, Alexander, and Pennekamp}]{Shen2023}
Shen C, Lemmen K, Alexander JM, Pennekamp F.
\newblock Connecting higher-order interactions with ecological stability in
  experimental aquatic food webs.
\newblock {\em Ecology and Evolution\/} {\bf 13} (2023) e10502.
\newblock \doi{10.1002/ece3.10502}.

\bibitem[{Li et~al.(2026)Li, Xiao, Jiang, Wright, Mayfield, Godoy
  et~al.}]{Li2026}
Li Y, Xiao J, Jiang Y, Wright SJ, Mayfield MM, Godoy O, et~al.
\newblock Higher-order interactions enhance the latitudinal tree diversity
  gradient.
\newblock {\em Nature\/} {\bf 653} (2026) 433--438.
\newblock \doi{10.1038/s41586-026-10434-6}.

\bibitem[{Cui et~al.(2024)Cui, III, and
  Mehta}]{cui2024leshoucheslecturescommunity}
Cui W, III RM, Mehta P.
\newblock Les houches lectures on community ecology: From niche theory to
  statistical mechanics  (2024).

\bibitem[{Calleja-Solanas et~al.(2026)Calleja-Solanas, Lamata-Ot{\'i}n,
  G{\'o}mez-Ambrosi, G{\'o}mez-Garde{\~n}es, and Meloni}]{CallejaSolanas2026}
Calleja-Solanas V, Lamata-Ot{\'i}n S, G{\'o}mez-Ambrosi C,
  G{\'o}mez-Garde{\~n}es J, Meloni S.
\newblock Higher-order interactions in ecology can be hidden in plain sight.
\newblock {\em arXiv preprint arXiv:2605.06301\/}  (2026).

\bibitem[{Guo et~al.(2021)Guo, Jia, Sendi{\~n}a-Nadal, Zhang, Wang, Li
  et~al.}]{Guo2021}
Guo H, Jia D, Sendi{\~n}a-Nadal I, Zhang M, Wang Z, Li X, et~al.
\newblock Evolutionary games on simplicial complexes.
\newblock {\em Chaos, Solitons \& Fractals\/} {\bf 150} (2021) 111103.
\newblock \doi{10.1016/j.chaos.2021.111103}.

\bibitem[{Civilini et~al.(2024)Civilini, Sadekar, Battiston,
  G{\'o}mez-Garde{\~n}es, and Latora}]{Civilini2024}
Civilini A, Sadekar O, Battiston F, G{\'o}mez-Garde{\~n}es J, Latora V.
\newblock Explosive cooperation in social dilemmas on higher-order networks.
\newblock {\em Physical Review Letters\/} {\bf 132} (2024) 167401.
\newblock \doi{10.1103/PhysRevLett.132.167401}.

\bibitem[{Nowak and May(1992)}]{NowakMay1992}
Nowak MA, May RM.
\newblock Evolutionary games and spatial chaos.
\newblock {\em Nature\/} {\bf 359} (1992) 826--829.
\newblock \doi{10.1038/359826a0}.

\bibitem[{Tka{\v{c}}ik et~al.(2014)Tka{\v{c}}ik, Marre, Amodei, Schneidman,
  Bialek, and Berry}]{Tkacik2014}
Tka{\v{c}}ik G, Marre O, Amodei D, Schneidman E, Bialek W, Berry MJ.
\newblock Searching for collective behavior in a large network of sensory
  neurons.
\newblock {\em PLoS Computational Biology\/} {\bf 10} (2014) e1003408.
\newblock \doi{10.1371/journal.pcbi.1003408}.

\bibitem[{Noonan and Lambiotte(2021)}]{Noonan2021}
Noonan J, Lambiotte R.
\newblock Dynamics of majority rule on hypergraphs.
\newblock {\em Physical Review E\/} {\bf 104} (2021) 024316.
\newblock \doi{10.1103/PhysRevE.104.024316}.

\bibitem[{Carletti et~al.(2020)Carletti, Battiston, Cencetti, and
  Fanelli}]{Carletti2020}
Carletti T, Battiston F, Cencetti G, Fanelli D.
\newblock Random walks on hypergraphs.
\newblock {\em Physical Review E\/} {\bf 101} (2020) 022308.
\newblock \doi{10.1103/PhysRevE.101.022308}.

\bibitem[{Castellani and Cavagna(2005)}]{CastellaniCavagna2005}
Castellani T, Cavagna A.
\newblock Spin-glass theory for pedestrians.
\newblock {\em Journal of Statistical Mechanics: Theory and Experiment\/} {\bf
  2005} (2005) P05012.
\newblock \doi{10.1088/1742-5468/2005/05/P05012}.

\bibitem[{M{\'e}zard and Montanari(2009)}]{MezardMontanari2009}
M{\'e}zard M, Montanari A.
\newblock {\em Information, Physics, and Computation\/}.
\newblock Oxford Graduate Texts (Oxford University Press) (2009).
\newblock \doi{10.1093/acprof:oso/9780198570837.001.0001}.

\bibitem[{Gardner(1985)}]{Gardner1985}
Gardner E.
\newblock Spin glasses with p-spin interactions.
\newblock {\em Nuclear Physics B\/} {\bf 257} (1985) 747--765.
\newblock \doi{10.1016/0550-3213(85)90374-8}.

\bibitem[{Gross et~al.(1985)Gross, Kanter, and Sompolinsky}]{Gross1985}
Gross DJ, Kanter I, Sompolinsky H.
\newblock Mean-field theory of the potts glass.
\newblock {\em Physical Review Letters\/} {\bf 55} (1985) 304.
\newblock \doi{10.1103/PhysRevLett.55.304}.

\bibitem[{Angelini and Biroli(2014)}]{AngeliniBiroli2014}
Angelini MC, Biroli G.
\newblock Super-{P}otts glass: A disordered model for glass-forming liquids.
\newblock {\em Physical Review B\/} {\bf 90} (2014) 220201.
\newblock \doi{10.1103/PhysRevB.90.220201}.

\bibitem[{Ackley et~al.(1985)Ackley, Hinton, and Sejnowski}]{Ackley1985}
Ackley DH, Hinton GE, Sejnowski TJ.
\newblock A learning algorithm for {B}oltzmann machines.
\newblock {\em Cognitive Science\/} {\bf 9} (1985) 147--169.
\newblock \doi{10.1016/S0364-0213(85)80012-4}.

\bibitem[{Le~Roux and Bengio(2008)}]{LeRouxBengio2008}
Le~Roux N, Bengio Y.
\newblock Representational power of restricted {B}oltzmann machines and deep
  belief networks.
\newblock {\em Neural Computation\/} {\bf 20} (2008) 1631--1649.
\newblock \doi{10.1162/neco.2008.04-07-510}.

\bibitem[{Landry and Restrepo(2020)}]{LandryRestrepo2020}
Landry NW, Restrepo JG.
\newblock The effect of heterogeneity on hypergraph contagion models.
\newblock {\em Chaos\/} {\bf 30} (2020) 103117.
\newblock \doi{10.1063/5.0020034}.

\bibitem[{Ferraz~de Arruda et~al.(2020)Ferraz~de Arruda, Petri, and
  Moreno}]{deArruda2020}
Ferraz~de Arruda G, Petri G, Moreno Y.
\newblock Social contagion models on hypergraphs.
\newblock {\em Physical Review Research\/} {\bf 2} (2020) 023032.
\newblock \doi{10.1103/PhysRevResearch.2.023032}.

\bibitem[{St-Onge et~al.(2021)St-Onge, Thibeault, Allard, Dub{\'e}, and
  H{\'e}bert-Dufresne}]{StOnge2021}
St-Onge G, Thibeault V, Allard A, Dub{\'e} LJ, H{\'e}bert-Dufresne L.
\newblock Universal nonlinear infection kernel from heterogeneous exposure on
  higher-order networks.
\newblock {\em Physical Review Letters\/} {\bf 127} (2021) 158301.
\newblock \doi{10.1103/PhysRevLett.127.158301}.

\bibitem[{Sun and Bianconi(2021)}]{SunBianconi2021}
Sun H, Bianconi G.
\newblock Higher-order percolation processes on multiplex hypergraphs.
\newblock {\em Physical Review E\/} {\bf 104} (2021) 034306.
\newblock \doi{10.1103/PhysRevE.104.034306}.

\bibitem[{P{\'e}rez-Reche et~al.(2011)P{\'e}rez-Reche, Ludlam, Taraskin, and
  Gilligan}]{PerezRecheLudlam2011}
P{\'e}rez-Reche FJ, Ludlam JJ, Taraskin SN, Gilligan CA.
\newblock Synergy in spreading processes: From exploitative to explorative
  foraging strategies.
\newblock {\em Physical Review Letters\/} {\bf 106} (2011) 218701.
\newblock \doi{10.1103/PhysRevLett.106.218701}.

\bibitem[{G{\'o}mez-Garde{\~n}es et~al.(2016)G{\'o}mez-Garde{\~n}es, Lotero,
  Taraskin, and P{\'e}rez-Reche}]{GomezGardenes2016}
G{\'o}mez-Garde{\~n}es J, Lotero L, Taraskin SN, P{\'e}rez-Reche FJ.
\newblock Explosive contagion in networks.
\newblock {\em Scientific Reports\/} {\bf 6} (2016) 19767.
\newblock \doi{10.1038/srep19767}.

\bibitem[{Taraskin and P{\'e}rez-Reche(2019)}]{Taraskin2019}
Taraskin SN, P{\'e}rez-Reche FJ.
\newblock Bifurcations in synergistic epidemics on random regular graphs.
\newblock {\em arXiv preprint arXiv:1809.05575\/}  (2019).

\bibitem[{P{\'e}rez-Reche and Taraskin(2025)}]{PerezReche2025}
P{\'e}rez-Reche FJ, Taraskin SN.
\newblock Multistability and high codimension bifurcations in synergistic
  epidemics on heterogeneous networks.
\newblock {\em Physical Review E\/} {\bf 112} (2025) L032301.
\newblock \doi{10.1103/pnvl-zkq7}.

\bibitem[{Granovetter(1978)}]{Granovetter1978}
Granovetter M.
\newblock Threshold models of collective behavior.
\newblock {\em American Journal of Sociology\/} {\bf 83} (1978) 1420--1443.
\newblock \doi{10.1086/226707}.

\bibitem[{Watts(2002)}]{Watts2002}
Watts DJ.
\newblock A simple model of global cascades on random networks.
\newblock {\em Proceedings of the National Academy of Sciences\/} {\bf 99}
  (2002) 5766--5771.
\newblock \doi{10.1073/pnas.082090499}.

\bibitem[{Dodds and Watts(2004)}]{DoddsWatts2004}
Dodds PS, Watts DJ.
\newblock Universal behavior in a generalized model of contagion.
\newblock {\em Physical Review Letters\/} {\bf 92} (2004) 218701.
\newblock \doi{10.1103/PhysRevLett.92.218701}.

\bibitem[{Centola et~al.(2007)Centola, Egu{\'i}luz, and
  Macy}]{CentolaEguiluzMacy2007}
Centola D, Egu{\'i}luz VM, Macy MW.
\newblock Cascade dynamics of complex propagation.
\newblock {\em Physica A: Statistical Mechanics and its Applications\/} {\bf
  374} (2007) 449--456.
\newblock \doi{10.1016/j.physa.2006.06.018}.

\bibitem[{Meloni et~al.(2026)Meloni, Gabrielli, and Villegas}]{Meloni2026}
Meloni S, Gabrielli A, Villegas P.
\newblock Higher-order contagion processes in 3.99 dimensions.
\newblock {\em Physical Review E\/} {\bf 113} (2026).
\newblock \doi{10.1103/98rn-9x9q}.

\bibitem[{Sun et~al.(2023)Sun, Radicchi, Kurths, and
  Bianconi}]{SunRadicchi2023}
Sun H, Radicchi F, Kurths J, Bianconi G.
\newblock The dynamic nature of percolation on networks with triadic
  interactions.
\newblock {\em Nature Communications\/} {\bf 14} (2023) 1308.
\newblock \doi{10.1038/s41467-023-37019-5}.

\bibitem[{Zwanzig(1961)}]{Zwanzig1961}
Zwanzig R.
\newblock Memory effects in irreversible thermodynamics.
\newblock {\em Physical Review\/} {\bf 124} (1961) 983.
\newblock \doi{10.1103/PhysRev.124.983}.

\bibitem[{Mori(1965)}]{Mori1965}
Mori H.
\newblock Transport, collective motion, and brownian motion.
\newblock {\em Progress of Theoretical Physics\/} {\bf 33} (1965) 423--455.
\newblock \doi{10.1143/PTP.33.423}.

\bibitem[{Chorin et~al.(2000)Chorin, Hald, and Kupferman}]{Chorin2000}
Chorin AJ, Hald OH, Kupferman R.
\newblock Optimal prediction and the mori-zwanzig representation of
  irreversible processes.
\newblock {\em Proceedings of the National Academy of Sciences of the United
  States of America\/} {\bf 97} (2000) 2968--2973.
\newblock
  \doi{10.1073/PNAS.97.7.2968;WEBSITE:WEBSITE:PNAS-SITE;WGROUP:STRING:PUBLICATION}.

\bibitem[{Cover and Thomas(2005)}]{Cover-Thomas_InformationTheoryBook}
Cover TM, Thomas JA.
\newblock {\em Elements of information theory\/} (Wiley-Interscience) (2005).
\newblock \doi{10.1002/047174882X}.

\bibitem[{Schreiber(2000)}]{Schreiber2000}
Schreiber T.
\newblock Measuring information transfer.
\newblock {\em Physical Review Letters\/} {\bf 85} (2000) 461--464.
\newblock \doi{10.1103/PhysRevLett.85.461}.

\bibitem[{Williams and Beer(2010)}]{WilliamsBeer2010}
Williams PL, Beer RD.
\newblock Nonnegative decomposition of multivariate information.
\newblock {\em arXiv preprint arXiv:1004.2515\/}  (2010).

\bibitem[{Bertschinger et~al.(2014)Bertschinger, Rauh, Olbrich, Jost, and
  Ay}]{Bertschinger2014}
Bertschinger N, Rauh J, Olbrich E, Jost J, Ay N.
\newblock Quantifying unique information.
\newblock {\em Entropy\/} {\bf 16} (2014) 2161--2183.
\newblock \doi{10.3390/e16042161}.

\bibitem[{Gutknecht et~al.(2021)Gutknecht, Wibral, and Makkeh}]{Gutknecht2021}
Gutknecht AJ, Wibral M, Makkeh A.
\newblock Bits and pieces: understanding information decomposition from
  part-whole relationships and formal logic.
\newblock {\em Proceedings of the Royal Society A\/} {\bf 477} (2021) 20210110.
\newblock \doi{10.1098/rspa.2021.0110}.

\bibitem[{Watanabe(1960)}]{Watanabe1960}
Watanabe S.
\newblock Information theoretical analysis of multivariate correlation.
\newblock {\em IBM Journal of Research and Development\/} {\bf 4} (1960)
  66--82.
\newblock \doi{10.1147/rd.41.0066}.

\bibitem[{Han(1978)}]{Han1978}
Han TS.
\newblock Nonnegative entropy measures of multivariate symmetric correlations.
\newblock {\em Information and Control\/} {\bf 36} (1978) 133--156.
\newblock \doi{10.1016/S0019-9958(78)90275-9}.

\bibitem[{Scagliarini et~al.(2023)Scagliarini, Nuzzi, Antonacci, Faes, Rosas,
  Marinazzo et~al.}]{Scagliarini2023}
Scagliarini T, Nuzzi D, Antonacci Y, Faes L, Rosas FE, Marinazzo D, et~al.
\newblock Gradients of o-information highlight synergy and redundancy in
  physiological applications.
\newblock {\em Frontiers in Network Physiology\/} {\bf 3} (2023) 1335808.
\newblock \doi{10.3389/fnetp.2023.1335808}.

\bibitem[{Stramaglia et~al.(2021)Stramaglia, Scagliarini, Daniels, and
  Marinazzo}]{Stramaglia2021}
Stramaglia S, Scagliarini T, Daniels BC, Marinazzo D.
\newblock Quantifying dynamical high-order interdependencies from the
  o-information: an application to neural spiking dynamics.
\newblock {\em Frontiers in Physiology\/} {\bf 11} (2021) 595736.
\newblock \doi{10.3389/fphys.2020.595736}.

\bibitem[{Matsuda(2000)}]{Matsuda2000}
Matsuda H.
\newblock Physical nature of higher-order mutual information: Intrinsic
  correlations and frustration.
\newblock {\em Physical Review E\/} {\bf 62} (2000) 3096.
\newblock \doi{10.1103/PhysRevE.62.3096}.

\bibitem[{Schneidman et~al.(2003)Schneidman, Still, Berry, and
  Bialek}]{Schneidman2003}
Schneidman E, Still S, Berry MJ, Bialek W.
\newblock Network information and connected correlations.
\newblock {\em Physical Review Letters\/} {\bf 91} (2003) 238701.
\newblock \doi{10.1103/PhysRevLett.91.238701}.

\bibitem[{Jaynes(1957)}]{Jaynes1957}
Jaynes ET.
\newblock Information theory and statistical mechanics.
\newblock {\em Physical Review\/} {\bf 106} (1957) 620--630.
\newblock \doi{10.1103/PhysRev.106.620}.

\bibitem[{Ganmor et~al.(2011)Ganmor, Segev, and Schneidman}]{Ganmor2011}
Ganmor E, Segev R, Schneidman E.
\newblock Sparse low-order interaction network underlies a highly correlated
  and learnable neural population code.
\newblock {\em Proceedings of the National Academy of Sciences\/} {\bf 108}
  (2011) 9679--9684.
\newblock \doi{10.1073/pnas.1019641108}.

\bibitem[{Roudi et~al.(2009)Roudi, Nirenberg, and Latham}]{Roudi2009}
Roudi Y, Nirenberg S, Latham PE.
\newblock Pairwise maximum entropy models for studying large biological
  systems: when they can work and when they can't.
\newblock {\em PLoS Computational Biology\/} {\bf 5} (2009) e1000380.
\newblock \doi{10.1371/journal.pcbi.1000380}.

\bibitem[{Santoro et~al.(2023)Santoro, Battiston, Petri, and
  Amico}]{Santoro2023}
Santoro A, Battiston F, Petri G, Amico E.
\newblock Higher-order organization of multivariate time series.
\newblock {\em Nature Physics\/} {\bf 19} (2023) 221--229.
\newblock \doi{10.1038/s41567-022-01852-0}.

\bibitem[{Kawatsu et~al.(2021)Kawatsu, Ushio, van Veen, and
  Kondoh}]{Kawatsu2021}
Kawatsu K, Ushio M, van Veen FJF, Kondoh M.
\newblock Are networks of trophic interactions sufficient for understanding the
  dynamics of multi-trophic communities? {A}nalysis of a tri-trophic insect
  food-web time-series.
\newblock {\em Ecology Letters\/} {\bf 24} (2021) 543--552.
\newblock \doi{10.1111/ele.13672}.

\bibitem[{Kawatsu(2024)}]{Kawatsu2024}
Kawatsu K.
\newblock Local-manifold-distance-based regression: an estimation method for
  quantifying dynamic biological interactions with empirical time series.
\newblock {\em Royal Society Open Science\/} {\bf 11} (2024) 231795.
\newblock \doi{10.1098/rsos.231795}.

\bibitem[{Malizia et~al.(2024)Malizia, Corso, Gambuzza, Russo, Latora, and
  Frasca}]{Malizia2024}
Malizia F, Corso A, Gambuzza LV, Russo G, Latora V, Frasca M.
\newblock Reconstructing higher-order interactions in coupled dynamical
  systems.
\newblock {\em Nature Communications\/} {\bf 15} (2024) 5184.
\newblock \doi{10.1038/s41467-024-49278-x}.

\bibitem[{Wang et~al.(2022)Wang, Ma, Chen, Lai, and Zhang}]{Wang2022}
Wang H, Ma C, Chen HS, Lai YC, Zhang HF.
\newblock Full reconstruction of simplicial complexes from binary contagion and
  ising data.
\newblock {\em Nature Communications\/} {\bf 13} (2022) 3043.
\newblock \doi{10.1038/s41467-022-30706-9}.

\bibitem[{Tang et~al.(2026)Tang, Srikrishnan, and Kulik}]{Tang2026}
Tang K, Srikrishnan V, Kulik J.
\newblock Bayesian hypergraph inference from scarce and noisy dynamical
  observations.
\newblock {\em arXiv preprint arXiv:2605.04218\/}  (2026).

\bibitem[{Young et~al.(2021)Young, Petri, and Peixoto}]{YoungPetriPeixoto2021}
Young JG, Petri G, Peixoto TP.
\newblock Hypergraph reconstruction from network data.
\newblock {\em Communications Physics\/} {\bf 4} (2021) 135.
\newblock \doi{10.1038/s42005-021-00637-w}.

\bibitem[{Yao et~al.(2021)Yao, Chen, Evans, and Christensen}]{Yao2021}
Yao Q, Chen B, Evans TS, Christensen K.
\newblock Higher-order temporal network effects through triplet evolution.
\newblock {\em Scientific Reports\/} {\bf 11} (2021) 15419.
\newblock \doi{10.1038/s41598-021-94389-w}.

\bibitem[{Musciotto et~al.(2021)Musciotto, Battiston, and
  Mantegna}]{Musciotto2021}
Musciotto F, Battiston F, Mantegna RN.
\newblock Detecting informative higher-order interactions in statistically
  validated hypergraphs.
\newblock {\em Communications Physics\/} {\bf 4} (2021) 1--9.
\newblock \doi{10.1038/s42005-021-00710-4}.

\bibitem[{Newman(2018)}]{Newman2018}
Newman ME.
\newblock Network structure from rich but noisy data.
\newblock {\em Nature Physics\/} {\bf 14} (2018) 542--545.
\newblock \doi{10.1038/s41567-018-0076-1}.

\bibitem[{Lizotte et~al.(2023)Lizotte, Young, and Allard}]{Lizotte2023}
Lizotte S, Young JG, Allard A.
\newblock Hypergraph reconstruction from uncertain pairwise observations.
\newblock {\em Scientific Reports\/} {\bf 13} (2023) 21364.
\newblock \doi{10.1038/s41598-023-48081-w}.

\bibitem[{Monroy(2026{\natexlab{a}})}]{Monroy2026a}
Monroy F.
\newblock Biological function as quantised organised action.
\newblock {\em BioSystems\/} {\bf 268} (2026{\natexlab{a}}) 105922.
\newblock \doi{10.1016/J.BIOSYSTEMS.2026.105922}.

\bibitem[{Monroy(2026{\natexlab{b}})}]{Monroy2026b}
Monroy F.
\newblock Hamilton-jacobi theory of quantized biological function.
\newblock {\em APS Open Science\/} {\bf 1} (2026{\natexlab{b}}) 000140.
\newblock \doi{10.1103/x2fd-pjlb}.

\bibitem[{LaRock and Lambiotte(2025)}]{LaRockLambiotte2025}
LaRock T, Lambiotte R.
\newblock Exploring the non-uniqueness of node co-occurrence matrices of
  hypergraphs.
\newblock {\em arXiv preprint arXiv:2506.01479\/}  (2025).

\bibitem[{Villegas and Meloni(2026)}]{Villegas2026}
Villegas P, Meloni S.
\newblock Is higher-order physics different?
\newblock {\em arXiv preprint arXiv:2609.20759\/}  (2026).

\bibitem[{Villaverde(2019)}]{Villaverde2019}
Villaverde AF.
\newblock Observability and structural identifiability of nonlinear biological
  systems.
\newblock {\em Complexity\/} {\bf 2019} (2019) 8497093.
\newblock \doi{10.1155/2019/8497093}.

\end{thebibliography}
\end{document}